\documentclass[aps,floats,twocolumn,prb,showpacs,10pt,superscriptaddress,nofootinbib]{revtex4}
\usepackage{graphicx}
\usepackage{color}
\usepackage{upgreek}
\usepackage{setspace}
\usepackage{amssymb}
\usepackage{amsmath}
\usepackage{dsfont}
\usepackage{fancyhdr}
\usepackage{array}
\usepackage{multirow}
\usepackage{booktabs}
\usepackage{diagbox}
\usepackage{hhline}
\usepackage{oubraces}
\usepackage{hyperref}

\allowdisplaybreaks[4]

\newcolumntype{M}[1]{>{\centering\arraybackslash}m{#1}}

\begin{document}
\title{Analytical investigation of singularities in two-particle irreducible vertex functions of the Hubbard atom}
\author{P. Thunstr\"om}
\affiliation{Uppsala University, Box 256, 751 05 Uppsala, Sweden}
\author{O. Gunnarsson}
\affiliation{Max-Planck-Institut f\"ur Festk\"orperforschung, Heisenbergstra\ss{}e 1, D-70569 Stuttgart, Germany}
\author{Sergio Ciuchi}
\affiliation{Department of Physical and Chemical Sciences, University of L'Aquila, Via Vetoio, I-67100 L'Aquila, Italy}
\affiliation{Consiglio Nazionale delle Ricerche (CNR-ISC) Via dei Taurini, I-00185 Rome, Italy}
\author{G. Rohringer}
\affiliation{Russian Quantum Center, Novaya street, 100, Skolkovo, Moscow region 143025, Russia}
\date{\today}

\begin{abstract}
Two-particle generalized susceptibilities and their irreducible vertex functions play a prominent role in the quantum many-body theory for correlated electron systems. They act as basic building blocks in the parquet formalism which provides a flexible scheme for the calculation of spectral and response functions. The irreducible vertices themselves have recently attracted increased attention as unexpected divergences in these functions have been identified. Remarkably, such singularities appear already for one of the simplest strongly interacting systems: the atomic limit of the half-filled Hubbard model (Hubbard atom). In this paper, we calculate the analytical expressions for {\em all} two-particle irreducible vertex functions of the Hubbard atom in {\em all} scattering channels as well as the fully irreducible two-particle vertices. We discuss their divergences and classify them by the eigenvalues and eigenvectors of the corresponding generalized susceptibilities. In order to establish a connection to the recently found multivaluedness of the exact self-energy functional $\Sigma[G]$, we show that already an approximation akin to iterated perturbation theory is sufficient to capture, qualitatively, the divergent structure of the vertex functions. Finally, we show that the localized divergences in the disordered binary mixture model are directly linked to a minimum in the single-particle Matsubara Green's function.
\end{abstract}

\maketitle

\let\n=\nu \let\o =\omega \let\s=\sigma


\section{Introduction}
\label{sec:intro}

One of the most successful tools for the theoretical description of strongly correlated electron systems are one- and many-particle Green's functions. They give access to a large number of experimentally measurable observables\cite{Abrikosov1975,Mahan2000}, such as the spectral function, the magnetic susceptibility, and the optical conductivity, which provide essential insights into the physics of realistic as well as model many-electron systems. Unfortunately, the one- and more-particle Green's functions are often very challenging to calculate in the presence of strong interactions between the particles. A breakthrough in this respect came with the advent of dynamical mean field theory (DMFT)\cite{Metzner1989,Georges1996} which maps the problem of correlated electrons on a lattice onto a single Anderson impurity model, i.e., a site which is embedded self-consistently into a dynamic bath. The Green's functions calculated within this approach include all purely local correlations in the system while nonlocal correlation effects are captured only on a mean field level. DMFT has been successfully exploited for describing many fascinating phenomena in correlated materials, such as the celebrated Mott metal-to-insulator transition\cite{Held2007}, the volume collapse in Ce\cite{Held2007}, magnetism in transition metals\cite{Lichtenstein2001,Hausoel2017}, and electronic entanglement in transition metal oxides\cite{Thunstrom2012}. The coherent potential approximation (CPA), serves a similar role in noninteracting disordered systems\cite{Soven1967}.

Self-consistent DMFT and CPA schemes are built around one-particle quantities, such as the local self-energy $\Sigma(\nu)$ and the lattice Green's function $G_{\mathbf{k}}(\nu)$\cite{Georges1996}. Nevertheless, in order to calculate nonlocal response functions (spin, charge, and pairing susceptibilities, optical conductivity, Hall coefficient, etc.) in the framework of DMFT and CPA, also the local two-particle generalized susceptibilities $\chi^{\nu\nu'\omega}_r$ are required as input to the Bethe-Salpeter (BS) equations\cite{Georges1996, Bickers2004, Rohringer2012}. In addition, the susceptibilities serve as building blocks for various diagrammatic extensions of DMFT\cite{Rohringer2018a} as the diagrammatic vertex approximation\cite{Toschi2007a,Rohringer2016} (D$\Gamma$A), the dual fermion (DF) method\cite{Rubtsov2008,Rubtsov2012}, and the quadruply-irreducible local expansion (QUADRILEX) \cite{Ayral2016a,Ayral2016} approach. 

Recently, a lot of attention has been directed towards the properties of the generalized susceptibilities as they,  unexpectedly, can become singular\cite{Rohringer2012,Schafer2013,Janis2014,Gunnarsson2016,Schafer2016,Chalupa2018,Vucicevic2018} in the intermediately-to-strongly correlated or disordered regime. The singular points are directly linked to eigenvalues of $\chi^{\nu\nu'\omega}_r$, for a fixed bosonic frequency $\omega$, that go from positive to negative when the interaction or disorder increases, and thus cross zero\cite{Rohringer2012,Schafer2016}. This implies that the corresponding local two-particle irreducible vertex ($\Gamma^{\nu\nu'\omega}_r$)\cite{Bickers2004,Rohringer2012,Rohringer2013a} as well as the local fully irreducible vertex $\Lambda^{\nu\nu'\omega}$ of the {\sl parquet formalism}\cite{Diatlov1957}, diverge at these points. Let us stress, that this behavior is {\em not} an artefact of the approximation introduced by the DMFT or CPA method as it is also observed in the case of infinite dimensions where these techniques provide an exact solution.  

The singularities of $\Gamma^{\nu\nu'\omega}_r$ and $\Lambda^{\nu\nu'\omega}$ are typically accompanied by crossings\cite{Kozik2015} of branches of the derivative of the Luttinger-Ward functional\cite{Baym1961,Luttinger1960} $\Phi[G]$, i.e., the self-energy functional $\Sigma[G]\!=\!\delta\Phi[G]/\delta G$.  It has been shown\cite{Gunnarsson2017} that such crossings lead to divergences in $\Gamma^{\nu\nu'\omega}_r$. In this respect, apart from being an indicator for the onset of (strong) correlations in a many-electron system, these divergences also limit the applicability of so-called bold diagrammatic Monte Carlo methods\cite{Prokofev1998,Prokofev2007} which sample the functional $\Sigma[G]$ by means of a Metropolis algorithm. In addition, the divergences of the {\em exact} fully irreducible vertex $\Lambda^{\nu\nu'\omega}$ restrict the applicability of diagrammatic extension of DMFT, which are based on this quantity\cite{Tam2013,Li2016}, and imply that the parquet approximation\cite{Bickers2004,Yang2009,Tam2013,Li2016,Kugler2018,Kugler2018a}, in which $\Lambda^{\nu\nu'\omega}$ is replaced by its lowest order (static) contribution $U$, must be applied with care. In simplified versions of the parquet equations (exploited, e.g., by the ladder approximations of D$\Gamma$A and DF\cite{Katanin2009,Rubtsov2008}), which neglect the mutual screening of the channels, the singularities of $\Gamma_r^{\nu\nu'\omega}$ and $\Lambda^{\nu\nu'\omega}$ can be in principle circumvented. However, it is currently unknown whether this approximation becomes less appropriate once the generalized susceptibilities start to become singular. 

A better understanding of the singular structure of the generalized susceptibilities $\chi^{\nu\nu'\omega}_r$ can help to address these questions and issues. Unfortunately, it is often difficult to draw definite conclusions based on purely numerical simulations, as numerical instabilities may hide asymptotic trends. Moreover, the quality of numerical results for $\Gamma_r^{\nu\nu'\omega}$ is limited by systematic truncation errors in the BS equations due to the restriction to a finite number of fermionic frequencies. Although recently methods have been suggested\cite{Kunes2011,Rohringer2012,Li2017,Wentzell2016,Tagliavini2018} to improve the treatment of the high-frequency asymptotic regime in the BS equations, many multi-orbital applications are still limited by the high numerical cost. Hence, analytical expressions for $\Gamma^{\nu\nu'\omega}_r$ and $\Lambda^{\nu\nu'\omega}$, and in particular for the eigenvalues and eigenvectors of $\chi^{\nu\nu'\omega}_r$, are highly desirable.

While for noninteracting disordered systems, such as the binary mixture (BM) or the Falicov-Kimball model (FKM), analytical results are indeed available\cite{Shvaika2000,Shvaika2001,Janis2014,Ribic2016,Ribic2016b}, no exact expressions are known for interacting systems. In this paper, we will fill this gap by presenting analytical results for $\Gamma^{\nu\nu'\omega}_r$ and $\Lambda^{\nu\nu'\omega}$ in a prototypical correlated system: the Hubbard model at half-filling in its atomic limit (referred to as Hubbard atom or simply AL in the following). Despite its very simple form this model exhibits complex two-particle correlations that display the above-mentioned singular structures.\cite{Schafer2013,Schafer2016,Gunnarsson2017} A subsequent analysis of the derived analytic formulas for $\Gamma^{\nu\nu'\omega}_r$, directly linked to the eigenvalues and eigenvectors of $\chi^{\nu\nu'\omega}_r$, provides therefore a step towards a deeper understanding of these divergences and the relation to the multivaluedness of the functional $\Sigma[G]$. Finally, the results of this paper can serve as a starting point for approximations to more complex strongly correlated systems.

The paper is organized as follows: In Sec.~\ref{sec:formalism} we recall the basic formalism of two-particle correlation functions for the Hubbard atom. In Sec.~\ref{sec:gamma} we present our analytical results for the two-particle irreducible vertices in all channels, while in Sec.~\ref{sec:eigen} we analyze their divergences in terms of the eigenvalues and eigenvectors of $\chi_r^{\nu\nu'\omega}$. In Sec.~\ref{sec:LW} we obtain a direct relation between the singularities of the irreducible vertex and the multivaluedness of the Luttinger-Ward functional by adopting an approximate expression for the (otherwise unknown) functional $\Sigma[G]$. In Sec.~\ref{sec:interpretation} we briefly outline possible interpretations of the vertex divergences and in Sec.~\ref{sec:1pGF} we allow ourselves to briefly speculate on the connection between the divergences of $\Gamma_r^{\nu\nu'\omega}$ and specific features of the single-particle Green's function. Finally, Sec.~\ref{sec:conclusions} is devoted to the conclusions and an outlook.

\section{Two-particle correlation functions for the AL}
\label{sec:formalism}

In the following, we will introduce the one- and two-particle Green's functions and the related generalized susceptibilities, as well as the two-particle irreducible vertex functions of a many-electron system. While in Sec.~\ref{subsec:gendef} the general definitions of these quantities are given, Sec.~\ref{subsec:defal} reviews the explicit expressions of the generalized susceptibilities of the half-filled Hubbard atom. As these subjects have been already discussed extensively in the literature, we will here just recapitulate the main points which are relevant for this work. For a more comprehensive discussion of the general two-particle formalism we refer the reader to the literature, in particular to Refs.~\onlinecite{Bickers2004,Rohringer2012,Rohringer2018a}.

\subsection{General definitions and formalism}
\label{subsec:gendef}

The Hubbard atom corresponds to an isolated $s$ orbital with an effective Coloumb interaction $U$ between the electrons. The chemical potential is set to $U/2$ to enforce particle-hole symmetry and half-filling, and in the absence of a magnetic field the system is also SU(2) symmetric with respect to the spin. Its Hamiltonian reads as
\begin{equation}
\label{equ:defhamilt}
\hat{\mathcal{H}}=U\hat{n}_{\uparrow}\hat{n}_{\downarrow}-\frac{U}{2}(\hat{n}_{\uparrow}+\hat{n}_{\downarrow}),
\end{equation}
where $\hat{n}_\sigma\!=\!\hat{c}^{\dagger}_\sigma\hat{c}_\sigma$ and $\hat{c}^{(\dagger)}_\sigma$ (creates) annihilates an electron with spin $\sigma=\uparrow,\downarrow$. The one-particle Green's function of this model is given by
\begin{equation}
\label{equ:1pgf}
G(\nu)=\frac{1}{i\nu-\frac{U^2}{4i\nu}},
\end{equation}
where $\nu\!=\!\frac{\pi}{\beta}(2n\!+\!1),\;n\!\in\!\mathds{Z}$, is a fermionic Matsubara frequency, and $\beta\!=\!1/T$ denotes the inverse temperature. Below, we will also use bosonic Matsubara frequencies which we denote as $\omega\!=\!\frac{\pi}{\beta}2m,\;m\!\in\!\mathds{Z}$.

The generalized susceptibility which is required for the calculation of the two-particle irreducible vertex functions is defined as
\begin{align}
\label{equ:defchi}
\chi^{\nu\nu'\omega}_{ph,\sigma\sigma'}&=\int\limits_{0}^{\beta}{d\tau_{1}d\tau_{2}d\tau_{3} \, e^{-i\nu\tau_{1}}e^{i(\nu+\omega)\tau_{2}}e^{-i(\nu'+\omega)\tau_{3}}} \nonumber\\
&\times \big[\big<T_{\tau}c_{\sigma}^{\dagger}(\tau_{1})c_{\sigma}(\tau_{2})c_{\sigma'}^{\dagger}(\tau_{3})c_{\sigma'}(0)\big> \nonumber\\
&- \big<T_{\tau}c_{\sigma}^{\dagger}(\tau_{1})c_{\sigma}(\tau_{2})\big>\big<T_{\tau}c_{\sigma'}^{\dagger}(\tau_{3})c_{\sigma'}(0)\big>\big].
\end{align}
Here, $T_{\tau}$ is the time-ordering operator and $\langle\ldots\rangle\!=\!\frac{1}{Z}\mbox{Tr}(e^{-\beta \hat{\mathcal{H}}}\ldots)$ denotes a thermal expectation value with $Z\!=\!\mbox{Tr}(e^{-\beta \hat{\mathcal{H}}})$. The assignment of the frequencies $\nu$, $\nu\!+\!\omega$ and $\nu'\!+\!\omega$ to the imaginary times $\tau_1$, $\tau_2$ and $\tau_3$, respectively, corresponds to the so-called particle-hole ($ph$) notation\cite{Rohringer2012}. Analogously, one can express the generalized susceptibility in the particle-particle ($pp$) representation which is obtained from the $ph$ one by a frequency shift, here defined as
\begin{equation}
\chi^{\nu\nu'\omega}_{{pp},\sigma\sigma'} \equiv \chi^{\nu\nu'(-\omega-\nu-\nu')}_{{ph},\sigma\sigma'}.
\end{equation}
The different physical interpretations of these two notations as particle-hole and particle-particle scattering amplitudes are discussed in detail in Ref.~\onlinecite{Rohringer2012}. Note that with respect to the latter, here we have defined (for convenience) the $pp$ notation with an additional minus sign for the bosonic frequency $\omega$.

{\renewcommand{\arraystretch}{2.0}
	\begin{table}[t!]
		\begin{tabular}{|c||c|c|c|c|}
			\hline
			& $d$ & $m$ & $s$ & $t$ \\
			\hhline{|=#=|=|=|=|}
			$A_r$ & $\frac{U}{2}\sqrt{3}$ & $i\frac{U}{2}$ & $0$ & $i\frac{U}{2}$ \\
			\hline
			$B_r$ & $\frac{U}{2}\sqrt{\frac{-1+3e^{\beta U/2}}{1+e^{\beta U/2}}}$ & $-\frac{U}{2}\sqrt{\frac{-1+3e^{-\beta U/2}}{1+e^{-\beta U/2}}}$ & $\frac{U}{2}\sqrt{\frac{-1+3e^{\beta U/2}}{1+e^{\beta U/2}}}$ & $0$ \\
			\hline
			$C_r^\omega$ & $\frac{\beta U}{2}\frac{\delta_{\omega 0}}{1+e^{\beta U/2}}$ & $-\frac{\beta U}{2}\frac{\delta_{\omega 0}}{1+e^{-\beta U/2}}$ & $\frac{\beta U}{2}\frac{\delta_{\omega 0}}{1+e^{\beta U/2}}$ & $0$ \\
			\hline
			$\mathcal{A}_0^r$ & $1$ & $1$ & $\frac{1}{2}$ & $-\frac{1}{2}$ \\
			\hline
			$\mathcal{B}_0^r$ & $1$ & $1$ & $\frac{1}{2}$ & $-\frac{1}{2}$ \\
			\hline
			$\mathcal{B}_1^r$ & $i$ & $1$ & $\frac{i}{\sqrt{2}}$ & $0$ \\
			\hline
			$\mathcal{B}_2^r$ & $1$ & $i$ & $\frac{1}{\sqrt{2}}$ & $0$ \\
			\hline
		\end{tabular}
		\caption{Prefactors $\mathcal{A}_0^r$, $\mathcal{B}_0^r$ ,$\mathcal{B}_1^r$ and $\mathcal{B}_2^r$ and constants $A_r$, $B_r$ and $C_r^\omega$ for the definition of the generalized susceptibilities $\chi_r^{\nu\nu'\omega}$ in the four channels $r=d,m,s,t$. $i$ denotes the imaginary unit which appears in $\mathcal{B}_1^r$ and $\mathcal{B}_2^r$ due to a negative sign of a factorized term [see last two summands in the definition of $\chi_r^{\nu\nu'\omega}$ in Eq.~(\ref{equ:chiAL})].}
		\label{tab:defprefactors}
	\end{table}}

For the SU(2) symmetric case considered here, it is convenient to use the spin-diagonalized versions of the generalized susceptibilities in the particle-hole as well as in the particle-particle notation. This corresponds to defining the $r=d\text{(ensity)},m\text{(agnetic)},s\text{(inglet)},t\text{(riplet)}$ generalized susceptibilities:\footnote{Note that the definitions for the singlet($s$) and triplet($t$) susceptibilities differ slightly from the corresponding ones given in Ref.~\onlinecite{Rohringer2012} [Eqs.~(B19) therein] in order to render the BS equations uniform in all channels.}
\begin{subequations}
	\label{equ:defsuscchannel}
	\begin{align}
	\label{equ:defsuscchanneld}
	&\chi_d^{\nu\nu'\omega}=\chi_{ph,\uparrow\uparrow}^{\nu\nu'\omega}+\chi_{ph,\uparrow\downarrow}^{\nu\nu'\omega},\\
	\label{equ:defsuscchannelm}
	&\chi_m^{\nu\nu'\omega}=\chi_{ph,\uparrow\uparrow}^{\nu\nu'\omega}-\chi_{ph,\uparrow\downarrow}^{\nu\nu'\omega},\\
	\label{equ:defsuscchannels}
	&\chi_s^{\nu\nu'\omega}=(-\chi_{pp,\uparrow\uparrow}^{\nu\nu'\omega}+2\chi_{pp,\uparrow\downarrow}^{\nu\nu'\omega}-2\chi_{0,pp}^{\nu\nu'\omega})/4,\\
	\label{equ:defsuscchannelt}
	&\chi_t^{\nu\nu'\omega}=(\chi_{pp,\uparrow\uparrow}^{\nu\nu'\omega}+2\chi_{0,pp}^{\nu\nu'\omega})/4,
	\end{align} 
\end{subequations}
where the bare particle-hole and particle-particle susceptibilities are defined as\footnote{A factor $\frac{1}{2}$ is included in the bare particle-particle susceptibility in order to take into account the indistinguishability of two particles which requires exactly such prefactor in the BS equation in the $pp$ channel (see Appendix B in Ref.~\onlinecite{Rohringer2012}). Hence, the transformation between the bare particle-hole and particle-particle susceptibility requires, in addition to the frequency shift discussed above for the full generalized susceptibilities, an additional factor 2.} 
\begin{subequations}
	\label{equ:defsuscbare}
	\begin{align}
	\label{equ:defsuscbareph}
	&\chi_{0,d/m}^{\nu\nu'\omega}=\chi_{0,ph}^{\nu\nu'\omega}=-\beta G(\nu)G(\nu+\omega)\delta_{\nu\nu'},\\
	\label{equ:defsuscbarepp}
	&\chi_{0,s/t}^{\nu\nu'\omega}=\chi_{0,pp}^{\nu\nu'\omega}=-\frac{\beta}{2} G(\nu)G(-\nu-\omega)\delta_{\nu\nu'}.
	\end{align}
\end{subequations}
The irreducible vertex functions $\Gamma_r^{\nu\nu'\omega}$ in all four channels ($d,m,s,t$) can be obtained from the generalized and bare susceptibilities by means of the BS equations\cite{Rohringer2012}
\begin{equation}
\label{equ:defBS}
\pm\chi_r^{\nu\nu'\omega}=\chi_{0,r}^{\nu\nu'\omega}-\frac{1}{\beta^2}\sum_{\nu_1\nu_2}\chi_{0,r}^{\nu\nu_1\omega}\Gamma_r^{\nu_1\nu_2\omega}\chi_r^{\nu_2\nu'\omega},
\end{equation}
where on the l.h.s. of this equation one has to take the plus sign for $\!r\!=d,m,t$ while for $r\!=\!s$ the minus sign has to be considered. Equation~(\ref{equ:defBS}) can be interpreted as a matrix equation for a fixed value of the bosonic frequency $\omega$ where the (discrete) fermionic frequencies $\nu$ and $\nu'$ represent the matrix indices. It can be solved for $\Gamma_r^{\nu\nu'\omega}$ by means of a matrix inversion with respect to $\nu$ and $\nu'$ which yields
\begin{equation}
\label{equ:BSinv}
\Gamma_r^{\nu\nu'\omega}=\beta^2(\chi_r^{-1}\mp\chi_{0,r}^{-1})^{\nu\nu'\omega},
\end{equation}
where for $r\!=\!d,m,t$ the minus sign and for $r\!=\!s$ the plus sign has to be taken.

As $\chi_{0,r}^{\nu\nu'\omega}$ is a diagonal matrix with only nonzero entries, its inverse is finite. Consequently, all divergences in $\Gamma_r^{\nu\nu'\omega}$ must originate from the inversion of $\chi_r^{\nu\nu'\omega}$ which will be analyzed explicitly in Sec.~\ref{sec:gamma}.

\subsection{Explicit expressions for the atomic limit}
\label{subsec:defal}

As shown in Appendix \ref{app:symmetries}, the particle-hole and spin SU(2) symmetry of the Hubbard atom entail the relation
\begin{equation}
	\chi_{\sigma\sigma'}^{\nu\nu'\omega} = \chi_{\sigma\sigma'}^{(-\nu-\omega)(-\nu'-\omega)\omega},\label{equ:chisymmetry}
\end{equation}
in both the particle-hole and particle-particle notations. The generalized susceptibilities can therefore be decomposed into a symmetric ($\chi_{r,\text{S}}^{\nu\nu'\omega}$) and an anti-symmetric ($\chi_{r,\text{A}}^{\nu\nu'\omega}$) part with respect to the transformation $\nu\!\rightarrow\!-\nu-\omega$. In the following, it is also convenient to explicitly keep track of the diagonal terms proportional to $\delta_{\nu\nu'}$ and $\delta_{\nu(-\nu'-\omega)}$. A particular feature of the atomic limit at half-filling is that its nondiagonal contributions can be factorized with respect to $\nu$ and $\nu'$. Hence, the general expression for the generalized susceptibility can be written in a unified form for all four channels as (see Refs.~\onlinecite{Hafermann2009b,Rohringer2012,Rohringer2013a})
\begin{widetext}
\begin{equation}
\label{equ:chiAL}
	\chi_r^{\nu\nu'\omega}=\overunderbraces{&\br{3}{\text{diagonal}}}
	{&a_{0,r}^{\nu\omega}[\delta_{\nu\nu'}-\delta_{\nu(-\nu'-\omega)}]&+&b_{0,r}^{\nu\omega}[\delta_{\nu\nu'}+\delta_{\nu(-\nu'-\omega)}]&+\sum_{l=1}^2b_{l,r}^{\nu\omega} b_{l,r}^{\nu'\omega}&.}{&\br{1}{\text{antisymmetric},\;\chi_{r,\text{A}}^{\nu\nu'\omega}} && \br{2}{\text{symmetric},\;\chi_{r,\text{S}}^{\nu\nu'\omega}}}
\end{equation}
\end{widetext}
The functions $a_{0,r}^{\nu\omega}$, $b_{0,r}^{\nu\omega}$, $b_{1,r}^{\nu\omega}$, and $b_{2,r}^{\nu\omega}$ are defined as
\begin{subequations}
\label{equ:defchisplit}
\begin{align}
\label{equ:defchisplit1}
	&a_{0,r}^{\nu\omega}=\mathcal{A}_0^r\frac{\beta}{2}\frac{[\nu(\nu+\omega)-A_r^2]}{[\nu^2+\frac{U^2}{4}][(\nu+\omega)^2+\frac{U^2}{4}]},\\[0.15cm]
	\label{equ:defchisplit2}
	&b_{0,r}^{\nu\omega}=\mathcal{B}_0^r\frac{\beta}{2}\frac{[\nu(\nu+\omega)-B_r^2]}{[\nu^2+\frac{U^2}{4}][(\nu+\omega)^2+\frac{U^2}{4}]},\\[0.15cm]
	\label{equ:defchisplit3}
	&b_{1,r}^{\nu\omega}=\mathcal{B}_1^r\frac{\sqrt{U(1-C_r^\omega)}[\nu(\nu+\omega)-D_r^\omega]}{[\nu^2+\frac{U^2}{4}][(\nu+\omega)^2+\frac{U^2}{4}]},\\[0.15cm]
	\label{equ:defchisplit4}
	&b_{2,r}^{\nu\omega}=\mathcal{B}_2^r\frac{\sqrt{\frac{U^3}{4}}\sqrt{\frac{U^2}{1-C_r^\omega}+\omega^2}}{[\nu^2+\frac{U^2}{4}][(\nu+\omega)^2+\frac{U^2}{4}]},
\end{align}
\end{subequations}
where $D_r^\omega$ is given by
\begin{equation}
\label{equ:defD}
	D_r^\omega=\frac{U^2}{4}\frac{1+C_r^\omega}{1-C_r^\omega}.
\end{equation}
The channel-dependent prefactors $\mathcal{A}_0^r$, $\mathcal{B}_0^r$, $\mathcal{B}_1^r$ and $\mathcal{B}_2^r$ as well as the channel-dependent constants $A_r$, $B_r$, and $C_r^\omega$ (which depends on $\omega$ only via $\delta_{\omega0}$) are given in Table~\ref{tab:defprefactors}. Here we note that the factor $\sqrt{3}$ in $A_d$ arises from the addition of the anti-symmetric parts of $\chi_{ph,\uparrow\uparrow}^{\nu\nu'\omega}$ and $\chi_{ph,\uparrow\downarrow}^{\nu\nu'\omega}$ in Eq.~(\ref{equ:defsuscchanneld}) and, hence, reflects the spin-$1/2$ nature of the particles (see also Sec.\ref{sec:LW}).

\section{Analytical calculation of \texorpdfstring{$\Gamma_r$}{Gamma}}
\label{sec:gamma}

Due to the special structure (\ref{equ:chiAL}) of $\chi_r^{\nu\nu'\omega}$, the matrix inversion in Eq.~(\ref{equ:BSinv}) can be performed analytically via the Woodbury matrix identity\cite{Hager1989}. In the following we will go through the explicit inversion procedure as the actual calculation highlights {\em how} the divergences of different types (localized vs. global) develop in $(\chi_r^{-1})^{\nu\nu'\omega}$. 

The defining equation of the inverse susceptibility, which we will denote by $\overline{\chi}_r^{\nu\nu'\omega}\!\equiv\!(\chi_r^{-1})^{\nu\nu'\omega}$ in the following, is given by
\begin{equation}
\label{equ:defchiinv}
\sum_{\nu_1}\chi_r^{\nu\nu_1\omega}\overline{\chi}_r^{\nu_1\nu'\omega}=\delta_{\nu\nu'}.
\end{equation}
Inserting the explicit expression for $\chi_r^{\nu\nu'\omega}$ in Eq.~(\ref{equ:chiAL}) into Eq.~(\ref{equ:defchiinv}) and the corresponding relation for $\chi_r^{(-\nu-\omega)\nu'\omega}$ yields the following two equations for $\overline{\chi}_r^{\nu\nu'\omega}$ and $\overline{\chi}_r^{(-\nu-\omega)\nu'\omega}$:
\begin{subequations}
	\label{equ:twobytwochiinv}
	\begin{align}
	\delta_{\nu\nu'} = & (a_{0,r}^{\nu\omega}+b_0^{\nu\omega})\overline{\chi}^{\nu\nu'\omega}+(-a_{0,r}^{\nu\omega}+b_{0,r}^{\nu\omega})\overline{\chi}^{(-\nu-\omega)\nu'\omega} \nonumber \\
	& + \sum_{l=1}^2 b_{l,r}^{\nu\omega}{Q_{l,r}^{\nu'\omega}}\label{equ:twobytwochiinv1}\\
	\delta_{(-\nu-\omega)\nu'} = & (a_{0,r}^{\nu\omega}+b_{0,r}^{\nu\omega})\overline{\chi}^{(-\nu-\omega)\nu'\omega}+(-a_{0,r}^{\nu\omega}+b_{0,r}^{\nu\omega})\overline{\chi}^{\nu\nu'\omega}\nonumber\\
	& + \sum_{l=1}^2 b_{l,r}^{\nu\omega}{Q_{l,r}^{\nu'\omega}},\label{equ:twobytwochiinv2}
	\end{align}
\end{subequations}
where we have defined
\begin{equation}
Q_{l,r}^{\nu'\omega} = \sum_{\nu_1}b_{l,r}^{\nu_1\omega}\overline{\chi}_r^{\nu_1\nu'\omega},\label{equ:qdef}
\end{equation}
and used the symmetries of the quantities $a_{0,r}^{\nu\omega}$, $b_{0,r}^{\nu\omega}$ and $b_{l,r}^{\nu\omega}$ under the (fermionic) frequency transformation $\nu\!\leftrightarrow\!-\nu\!-\!\omega$. The two Eqs.~(\ref{equ:twobytwochiinv}) can be combined to yield 
\begin{align}
\label{equ:chiinv}
\overline{\chi}_r^{\nu\nu'\omega}=&\frac{1}{4a_{0,r}^{\nu\omega}}[\delta_{\nu\nu'}-\delta_{\nu(-\nu'-\omega)}]+\frac{1}{4b_{0,r}^{\nu\omega}}[\delta_{\nu\nu'}+\delta_{\nu(-\nu'-\omega)}]\nonumber\\
&-\frac{1}{2b_{0,r}^{\nu\omega}}\sum_{l=1}^2b_{l,r}^{\nu\omega}Q_{l,r}^{\nu'\omega}.
\end{align}
The remaining task is now to determine $Q_{l,r}^{\nu'\omega}$. To this end we substitute Eq.~(\ref{equ:chiinv}) into Eq.~(\ref{equ:qdef}) to obtain a system of two linear equations for the two unknowns $Q_{1,r}^{\nu'\omega}$ and $Q_{2,r}^{\nu'\omega}$
\begin{subequations}
	\begin{align}
	\sum_{l=1}^2 M_{kl}^{r,\omega} Q_{l,r}^{\nu'\omega} = \frac{b_k^{\nu'\omega}}{2b_{0,r}^{\nu'\omega}},\label{equ:calcq1}\\
	M_{kl}^{r,\omega} = \delta_{kl} + \sum_\nu\frac{b_{k,r}^{\nu\omega}b_{l,r}^{\nu\omega}}{2b_{0,r}^{\nu\omega}}\label{equ:calcq12}.
	\end{align}
\end{subequations}
Equation~(\ref{equ:calcq1}) is solved straightforwardly by inverting the matrix $M_{kl}^{r,\omega}$ (see Appendix~\ref{app:evalnondiag}). The final expression for the inverse of the generalized susceptibility then reads as
\begin{align}
\label{equ:chibarfinal}
\overline{\chi}_r^{\nu\nu'\omega}=&\frac{1}{4a_{0,r}^{\nu\omega}}[\delta_{\nu\nu'}-\delta_{\nu(-\nu'-\omega)}]+\frac{1}{4b_{0,r}^{\nu\omega}}[\delta_{\nu\nu'}+\delta_{\nu(-\nu'-\omega)}] \nonumber\\
&-\frac{1}{4b_{0,r}^{\nu\omega}b_{0,r}^{\nu'\omega}}\sum_{k,l=1}^2b_{k,r}^{\nu\omega}\overline{M}_{kl}^{r,\omega}b_{l,r}^{\nu'\omega},
\end{align}
where $\overline{M}^{r,\omega}\!=\!(M^{r,\omega})^{-1}$ is the inverse of $M^{r,\omega}$. The matrix $M_{kl}^{r,\omega}$ in Eq.~(\ref{equ:calcq12}) can be obtained using the explicit definitions of $b_{0,r}^{\nu\omega}$, $b_{1,r}^{\nu\omega}$ and $b_{2,r}^{\nu\omega}$ [see Eqs.~(\ref{equ:defchisplit})]. The actual calculations involve lengthy, but analytically evaluable, sums over fermionic Matsubara frequencies which have been performed using {\em Mathematica}\cite{Wolfram2015}. The corresponding explicit calculations and results are reported in Appendix~\ref{app:matrixelements} as well as the Supplemental Material (SM)\cite{FootnoteSupplemental2018} where the {\em Mathematica} notebooks, which have been employed for this task, are given.

From Eq.~(\ref{equ:chibarfinal}) we can now easily derive the explicit analytical expressions for the irreducible vertex functions $\Gamma_r^{\nu\nu'\omega}$ by subtracting the inverse of $\chi_{0,r}^{\nu\nu'\omega}$ according to Eq.~(\ref{equ:BSinv}). This finally yields:
\begin{samepage}
\begin{widetext}
	\begin{align}
	\label{equ:gammafinal}
	\Gamma_r^{\nu\nu'\omega}=&\frac{\beta A_r^2}{2\mathcal{A}_0^r}\frac{\left(\nu^2+\frac{U^2}{4}\right)\left((\nu+\omega)^2+\frac{U^2}{4}\right)}{\left[\nu(\nu+\omega)-A_r^2\right]\left[\nu(\nu+\omega)\right]}[\delta_{\nu\nu'}-\delta_{\nu(-\nu'-\omega)}]+\frac{\beta B_r^2}{2\mathcal{B}_0^r}\frac{\left(\nu^2+\frac{U^2}{4}\right)\left((\nu+\omega)^2+\frac{U^2}{4}\right)}{\left[\nu(\nu+\omega)-B_r^2\right]\left[\nu(\nu+\omega)\right]}[\delta_{\nu\nu'}+\delta_{\nu(-\nu'-\omega)}]\nonumber\\[0.25cm]&-U\frac{\lvert \mathcal{B}_2^r\rvert^2}{(\mathcal{B}_0^r)^2}\frac{\frac{U^2}{4}\Bigl[\frac{U^2}{4}\left(\frac{4B_r^2}{U^2}+1\right)^2+\omega^2\Bigr]}{\frac{U\tan[\frac{\beta}{4}(\sqrt{4B_r^2+\omega^2}+\omega)]}{\sqrt{4B_r^2+\omega^2}}\pm1}\frac{1}{\nu(\nu+\omega)-B_r^2}\frac{1}{\nu'(\nu'+\omega)-B_r^2}-\left(\frac{\mathcal{B}_1^r}{\mathcal{B}_0^r}\right)^2U,
	\end{align}
\end{widetext}
\end{samepage}
where the plus sign in the denominator of the first term of the second line corresponds to $r\!=\!d,s$ while the minus sign has to be taken for $r\!=\!m$. The choice of sign does not affect the triplet vertex $\Gamma_t^{\nu\nu'\omega}$ since $\mathcal{B}_2^t\!=\!0$ (see Table~\ref{tab:defprefactors}). The terms in the first line of Eq.~(\ref{equ:gammafinal}) represent the diagonal contributions to the vertex irreducible in channel $r$ while the lower line  corresponds to the contributions which factorize with respect to the fermionic Matsubara frequencies $\nu$ and $\nu'$. 

\subsection{Divergences of \texorpdfstring{$\Gamma_r$}{Gamma}}
\label{subsec:divergchibar}

The explicit expression for the irreducible vertex $\Gamma_r^{\nu\nu'\omega}$ in Eq.~(\ref{equ:gammafinal}) allows us now to identify and classify {\em all} divergences of this function in all channels. Obviously, a singularity in $\Gamma_r^{\nu\nu'\omega}$ has to be expected when one of the denominators in Eq.~(\ref{equ:gammafinal}) vanishes. If this happens for a single frequency $\nu$ (and its crossing-symmetric counterpart $-\nu\!-\!\omega$), we are dealing with a localized divergence while the vanishing of a $\nu$-independent denominator gives rise to a global divergence (for a fixed value of $\omega$). A closer inspection of Eq.~(\ref{equ:gammafinal}) indicates three possible types of divergences:

(i) The denominator of the first summand in the first line of Eq.~(\ref{equ:gammafinal}) vanishes if
\begin{equation}
\label{equ:reddiv}
\nu(\nu+\omega)=A_r^2,
\end{equation}
which corresponds to $a_{0,r}^{\nu\omega}\!=\!0$ in Eq.~(\ref{equ:chibarfinal}). This condition gives clearly rise to a localized divergence in the $\nu,\nu'$ frequency space since, for a given value of $A_r$ and $\beta$, it can be fulfilled by only one Matsubara frequency $\nu^*$ and its symmetric conjugate $-\nu^*\!-\omega$. For the density, magnetic, and triplet channels ($r\!=\!d,m,t$), Eq.~(\ref{equ:reddiv}) can be fulfilled for specific values of $U$ if $\omega\!\ne\!0$. For the singlet channel, this is never possible at finite temperature since $A_s\!=\!0$ (see Table~\ref{tab:defprefactors}). For the special case of $\omega\!=\!0$, on the other hand, a solution to Eq.~(\ref{equ:reddiv}) can be found {\em only} for the density channel since $A_r$ is imaginary (and, hence, $A_r^2$ is negative) for the magnetic and triplet channels. This leads to localized divergences in $\Gamma_d^{\nu\nu'(\omega=0)}$ at frequencies $\nu^*\!=\!\pm\frac{U}{2}\sqrt{3}$ which have already been reported in Refs.~\onlinecite{Schafer2013,Rohringer2013a,Schafer2016}.

(ii) From the second and the third summands of $\Gamma_r^{\nu\nu'\omega}$ in Eq.~(\ref{equ:gammafinal}) we would expect the emergence of localized divergences if
\begin{equation}
\label{equ:red2div}
\nu(\nu+\omega)=B_r^2.
\end{equation}
However, the rule of l'H\^opital yields that the divergences in the two terms cancel. The analysis of the eigenvalues of $\chi_r^{\nu\nu'\omega}$ in the next section shows that the cancellation of these divergences is not accidental, but is directly linked to the shape of the eigenvectors.

\begin{figure}[t!]
	\centering
	\includegraphics[width=\columnwidth]{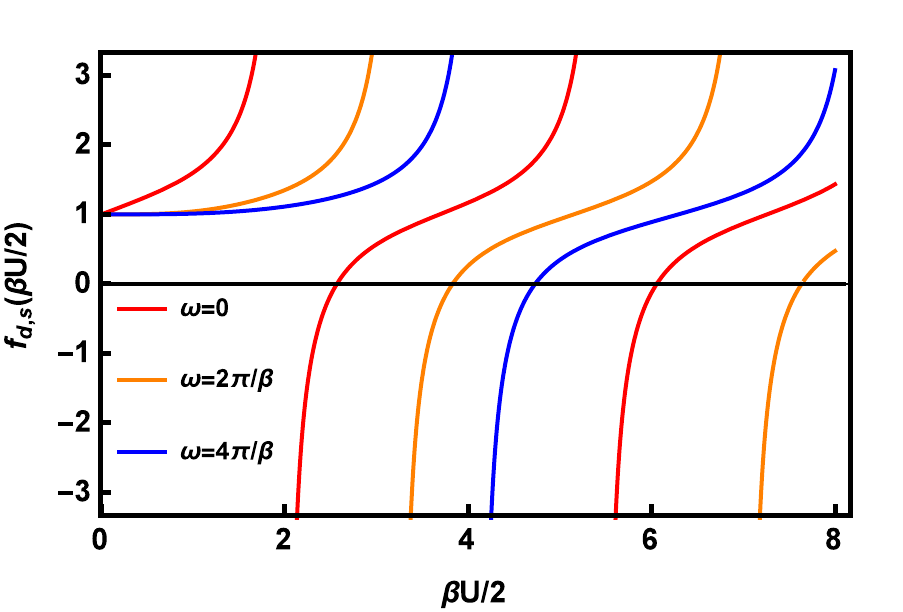}
	\includegraphics[width=\columnwidth]{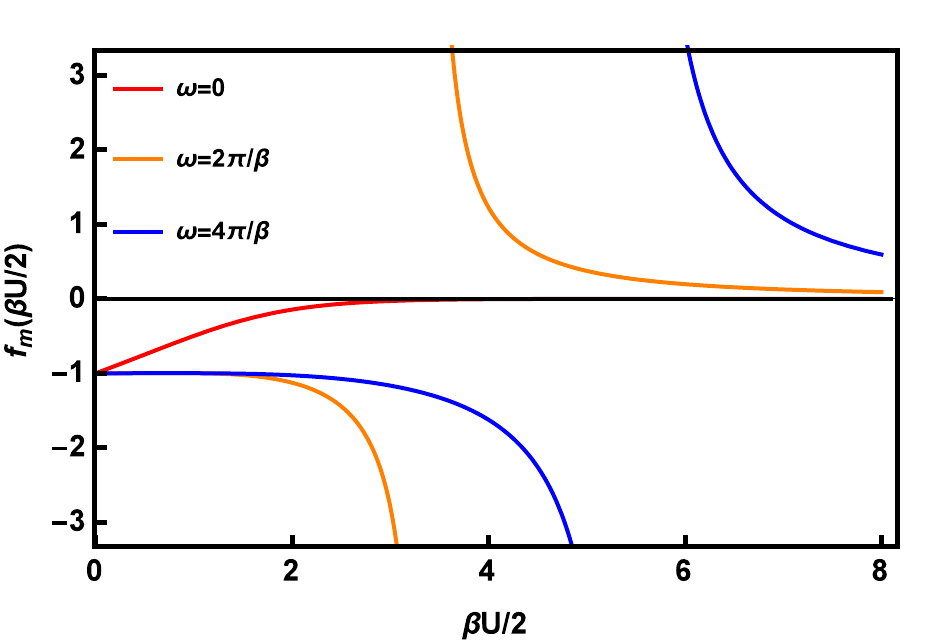}
	\caption{Function $f_{d,s}(\beta U/2)$ (upper panel) and $f_{m}(\beta U/2)$ (lower panel) for the first three bosonic Matsubara frequencies $\omega\!=\!0,2\pi/\beta,4\pi/\beta$. The crossings of the curves with the abscissa indicate a global divergence of the corresponding vertex.}
	\label{fig:orange_div_charge_singlet}
\end{figure}

(iii) Finally, $\Gamma_r^{\nu\nu'\omega}$ will diverge when the $\nu$-independent denominator in the second line of Eq.~(\ref{equ:gammafinal}) vanishes, i.e., if\footnote{Note that for $\omega\!=\!0$ this relation slightly differs from the corresponding one reported in Ref.~\onlinecite{Schafer2016} [see Eq.~(49) therein] for the density channel due to a typo in the latter: In fact, the term $\beta U$ on the left-hand side of Eq.~(49) in Ref.~\onlinecite{Schafer2016} has to be replaced by $1/(\beta U)$.}
\begin{align}
\label{equ:gammaglobdiv}
f_r(\beta U/2,\omega)\equiv\frac{U\tan[\frac{\beta}{4}(\sqrt{4B_r^2+\omega^2}+\omega)]}{\sqrt{4B_r^2+\omega^2}}\pm1=0.
\end{align}
where the upper (plus) sign has to be taken for the density and singlet ($r\!=\!d,s$) and the lower (minus) one for the magnetic ($r\!=\!m$) channel, respectively. For the triplet channel there is obviously no global divergence since the corresponding $\chi_t^{\nu\nu'\omega}$ consists of diagonal terms only. Equation~(\ref{equ:gammaglobdiv}) corresponds to a singularity of the matrix $M^{r,\omega}$ in Eq.~(\ref{equ:calcq12}), i.e., to the vanishing of the determinant of this matrix [see Eqs.~(\ref{equ:chibarfinal}) and (\ref{equ:det}) in Appendix~\ref{app:evalnondiag}]. Condition (\ref{equ:gammaglobdiv}) obviously gives rise to a {\em global} divergence of $\Gamma_r^{\nu\nu'\omega}$ since it is associated with a divergence of the prefactor in the second line of Eq.~(\ref{equ:gammafinal}) which does not depend on the fermionic Matsubara frequencies $\nu$ and $\nu'$.

Equation~(\ref{equ:gammaglobdiv}) represents a transcendental equation for the quantity $\frac{\beta U}{2}$ (cf. the definitions of $B_r$ in Table~\ref{tab:defprefactors}), which can be solved numerically. $f_r(\beta U/2,\omega)$ is plotted in Fig.~\ref{fig:orange_div_charge_singlet} in the density and singlet (upper panel) as well as for the magnetic channel (lower panel). A divergence occurs whenever a curve crosses $0$.

For the density and the singlet channels we find crossings of $f_{d,s}$ with the $x$ axis for each $\omega$ if $U>0$. The periodicity of the $\tan$ function in Eq.~(\ref{equ:gammaglobdiv}) then implies an {\em infinite} number of such global divergences, as predicted already in Ref.~\onlinecite{Schafer2016}. For the magnetic channel (lower panel in Fig.~\ref{fig:orange_div_charge_singlet}), on the contrary, $f_m(\beta U/2)\!\ne\!0$ for all values of $U>0$, $\beta$, and $\omega$. Hence, we conclude that there are no global divergences for $\Gamma_m^{\nu\nu'\omega}$ in the half-filled repulsive Hubbard atom. Let us mention that for $f_m$ no periodic behavior can be observed (for $U>0$) since the argument of the $\tan$ function becomes imaginary at $\frac{\beta U}{2}\!=\!\log[3+(\omega/U)^2]-\log[1-(\omega/U)^2]$ and, hence, the $\tan$ turns into $\tanh$. 

Let us give an estimate for the divergence condition in Eq.~(\ref{equ:gammaglobdiv}) for the density/singlet channel in the limit $\frac{\beta U}{2}\!\rightarrow\!\infty$ for $U > 0$ and a fixed $\omega$. A corresponding analysis has been already performed at $\omega = 0$ in a previous work\cite{Schafer2016} where, unfortunately, an incorrect result has been presented. We can rewrite Eq.~(\ref{equ:gammaglobdiv}) as
\begin{align}
\label{equ:globasymp}
-\frac{2\nu^* + \omega}{2U} & = \frac{1}{2U}\sqrt{4B_r^2 + \omega^2} - h(\omega),\\
h(\omega) & \equiv \frac{\pi}{\beta U}  +\frac{2}{\beta U} \arctan\big(\!\mp\frac{1}{U}\sqrt{4B_r^2 + \omega^2}\big),
\end{align}
where the fermionic frequency $\nu^*$ takes into account the infinitely many branches of the $\arctan$ function in $h(\omega)$. Taking the square of both sides\ of Eq.~(\ref{equ:globasymp})\footnote{Note that this operation does {\em not} introduce new solutions, since both sides of Eq.~(\ref{equ:globasymp}) are real valued and the frequency transformation $\nu^* \rightarrow -\nu^*-\omega$ can simply change the sign of the left-hand side.} leads to the condition 
\begin{equation}
\frac{\nu^*(\nu^* + \omega)}{U^2} = \frac{B_r^2}{U^2} - \frac{h(\omega)}{U}\sqrt{4B_r^2 + \omega^2} + h(\omega)^2,\label{equ:globasympsquare}
\end{equation}
for the occurrence of a global divergence. When $\frac{\beta U}{2} \!\rightarrow\!\infty$ (i.e., $T\rightarrow 0$) the value of $B_d=B_s$ reduces to $\frac{U}{2}\sqrt{3}$ and $h(\omega) \rightarrow 0$. Hence, Eq.~(\ref{equ:globasympsquare}) for the global divergences of $\Gamma_d^{\nu\nu'\omega}$ and $\Gamma_s^{\nu\nu'\omega}$ approaches the condition for the emergence of a localized divergence of type (i) in the charge channel [Eq.~(\ref{equ:reddiv})]. At fixed $U$, the loci of both types of singularities become infinitely dense for $T\rightarrow 0$ which makes $T=0$ a cluster point of singularities. 

Let us finally remark that Eq.~(\ref{equ:globasympsquare}) seems to define an energy scale $\nu^*$, as it has been observed for localized singularities [see Eq.~(\ref{equ:reddiv})]. The two situations are, however, different: While $\nu^*$ in the context of the localized singularities indeed defines an energy scale, i.e., a Matsubara frequency, at which a divergence occurs in $\Gamma_d^{\nu\nu'\omega}$, this is not true for the global singularities where the divergence takes place at all frequencies for all $T>0$. It is, hence, not clear how the (``artificial'') Matsubara frequency $\nu^*$ in Eq.~(\ref{equ:globasymp}) could be interpreted in the latter case. In this respect, it should be said that it is currently also unclear whether taking the limit $T\rightarrow 0$ yields the same result for $\Gamma_r^{\nu\nu'\omega}$ as the direct calculation at $T=0$. Moreover, at this point a spontaneous breaking of the SU(2) spin symmetry might occur which alters the formal structure of the BS equation as the spin-singlet ($r=d,s$) and spin-triplet ($r=m,t$) channels are not well defined any more. Whether this leads to a mitigation of singularities in the two-particle correlation functions is an interesting future research direction.  

\subsection{The fully irreducible vertex \texorpdfstring{$\Lambda_r$}{Lambda}}
\label{sec:lambda}

The analytical expression for $\Gamma_r^{\nu\nu'\omega}$ and $\chi_r^{\nu\nu'\omega}$ can be inserted into the parquet equations\cite{Diatlov1957,Janis2001,Bickers2004,Rohringer2012} to yield an analytic expression also for the fully irreducible vertex $\Lambda_r^{\nu\nu'\omega}$ in the atomic limit. In their SU(2) symmetric form, these equations can be written as

\begin{subequations}
\label{equs:parquet}
\begin{align}
\label{equ:parquetchannels}
\Lambda_d^{\nu\nu'\omega}&=\Gamma_d^{\nu\nu'\omega}
-\frac{1}{2}\Gamma_d^{\nu(\nu+\omega)(\nu'-\nu)}-\frac{3}{2}\Gamma_m^{\nu(\nu+\omega)(\nu'-\nu)}\notag\\
&\hspace{-0.9cm}+\frac{1}{2}\Gamma_s^{\nu\nu'(-\nu-\nu'-\omega)}+\frac{3}{2}\Gamma_t^{\nu\nu'(-\nu-\nu'-\omega)}-2F_d^{\nu\nu'\omega}
\\[0.3cm]
\Lambda_m^{\nu\nu'\omega}&=\Gamma_m^{\nu\nu'\omega}
-\frac{1}{2}\Gamma_d^{\nu(\nu+\omega)(\nu'-\nu)}+\frac{1}{2}\Gamma_m^{\nu(\nu+\omega)(\nu'-\nu)}\notag\\
&\hspace{-0.9cm}-\frac{1}{2}\Gamma_s^{\nu\nu'(-\nu-\nu'-\omega)}+\frac{1}{2}\Gamma_t^{\nu\nu'(-\nu-\nu'-\omega)}-2F_m^{\nu\nu'\omega}
\\[0.3cm]
\Lambda_s^{\nu\nu'\omega}&=\Gamma_s^{\nu\nu'\omega}
+\frac{1}{2}\Gamma_d^{\nu\nu'(-\nu-\nu'-\omega)}-\frac{3}{2}\Gamma_m^{\nu\nu'(-\nu-\nu'-\omega)}\notag\\
&\hspace{-0.9cm}+\frac{1}{2}\Gamma_d^{\nu(-\nu'-\omega)(\nu'-\nu)}-\frac{3}{2}\Gamma_m^{\nu(-\nu'-\omega)(\nu'-\nu)}-2F_s^{\nu\nu'\omega}
\\[0.3cm]
\Lambda_t^{\nu\nu'\omega}&=\Gamma_t^{\nu\nu'\omega}
+\frac{1}{2}\Gamma_d^{\nu\nu'(-\nu-\nu'-\omega)}+\frac{1}{2}\Gamma_m^{\nu\nu'(-\nu-\nu'-\omega)}\notag\\
&\hspace{-0.9cm}-\frac{1}{2}\Gamma_d^{\nu(-\nu'-\omega)(\nu'-\nu)}-\frac{1}{2}\Gamma_m^{\nu(-\nu'-\omega)(\nu'-\nu)}-2F_t^{\nu\nu'\omega},
\end{align}
\end{subequations}

where the full two-particle scattering amplitude $F_r^{\nu\nu'\omega}$ is given by
\begin{equation}
\label{equ:defF}
F_r^{\nu\nu'\omega}=-\frac{\chi_r^{\nu\nu'\omega}\mp\chi_{0,r}^{\nu\nu'\omega}}{\frac{1}{\beta^2}\sum_{\nu_1\nu_2}\chi_{0,r}^{\nu\nu_1\omega}\chi_{0,r}^{\nu_2\nu'\omega}}.
\end{equation}
Here, $\Lambda^{\nu\nu'\omega}_s$ and $\Lambda^{\nu\nu'\omega}_t$ as well as $F_s^{\nu\nu'\omega}$ and $F_t^{\nu\nu'\omega}$ are represented in particle-particle notation. Since $\Lambda_r^{\nu\nu'\omega}$ and $F_r^{\nu\nu'\omega}$ contain all fully two- and one-particle irreducible diagrams, respectively, the index $r$ refers only to the specific spin combination and the frequency notation in which the vertex is given rather than to a specific scattering channel. Hence, $\Lambda_s^{\nu\nu'\omega}$ and $\Lambda_t^{\nu\nu'\omega}$ can be expressed in terms of the $\Lambda_d^{\nu\nu'\omega}$ and $\Lambda_m^{\nu\nu'\omega}$:
\begin{equation}
\label{equ:parquetchanneldependence}
\begin{split}
&\Lambda_s^{\nu\nu'\omega}=\frac{1}{2}\Lambda_d^{\nu\nu(-\nu-\nu'-\omega)}-\frac{3}{2}\Lambda_m^{\nu\nu'(-\nu-\nu'-\omega)}\\
&\Lambda_t^{\nu\nu'\omega}=\frac{1}{2}\Lambda_d^{\nu\nu(-\nu-\nu'-\omega)}+\frac{1}{2}\Lambda_m^{\nu\nu'(-\nu-\nu'-\omega)},
\end{split}
\end{equation}
and the same relations hold for $F_s^{\nu\nu'\omega}$ and $F_t^{\nu\nu'\omega}$.

The fully irreducible vertices $\Lambda_r^{\nu\nu'\omega}$ can be now straightforwardly obtained by inserting the expressions for $\Gamma_r^{\nu\nu'\omega}$ in Eq.~(\ref{equ:gammafinal}) and $F_r^{\nu\nu'\omega}$ in Eq.~(\ref{equ:defF}) into Eqs.~(\ref{equs:parquet}). However, although $\Lambda_r^{\nu\nu'\omega}$ represents just a subset of the scattering processes found in $\Gamma_r^{\nu\nu'\omega}$, no appreciable simplification of the analytical expressions occur and, hence, for the final rather lengthy results we refer the reader to the SM\cite{FootnoteSupplemental2018} (c.f. the corresponding {\em Mathematica} notebook therein). In particular, the diverging terms of the different $\Gamma_r^{\nu\nu'\omega}$'s do {\em not} cancel out but remain in $\Lambda_r^{\nu\nu'\omega}$.

\section{Analysis of the eigenvectors}
\label{sec:eigen}

As shown in Sec.~\ref{subsec:gendef}, all divergences in $\Gamma_r^{\nu\nu'\omega}$ must originate from the inversion of $\chi_r^{\nu\nu'\omega}$. A divergence will therefore occur when one of the eigenvalues of the matrix $\chi_r^{\nu\nu'\omega}$ vanishes, and the diverging component will take the shape of the outer product of the corresponding eigenvectors. To gain a deeper understanding of the divergences in $\Gamma_r^{\nu\nu'\omega}$ we, hence, consider the eigenvalue equation
\begin{equation}
\label{equ:evequ}
\sum_{\nu'}\chi_r^{\nu\nu'\omega}V_r^{\nu'\omega}=\lambda_r^{\omega}V_r^{\nu\omega},
\end{equation}
where $\lambda_r^\omega$ denotes the eigenvalue and $V_r^{\nu\omega}$ is the corresponding eigenvector. 

In order to solve the eigenvalue equation~(\ref{equ:evequ}), let us first recall that the generalized susceptibility $\chi_r^{\nu\nu'\omega}$ can be decomposed into an anti-symmetric ($\chi_{r,\text{A}}^{\nu\nu'\omega}$) and a symmetric ($\chi_{r,\text{S}}^{\nu\nu'\omega}$) part [see Eq.~(\ref{equ:chiAL})] with respect to the transformation $\nu^{(\prime)}\!\rightarrow\!-\nu^{(\prime)}\!-\!\omega$. This observation implies that the eigenvector $V_r^{\nu\omega}$ is either antisymmetric ($V_{r,\text{A}}^{\nu\omega}$) or symmetric ($V_{r,\text{S}}^{\nu\omega}$), where $V_{r,\text{A}/\text{S}}^{(-\nu-\omega)\omega}\!=\!\mp V_{r,\text{A}/\text{S}}^{\nu\omega}$. Consequently, the eigenvalue problem splits into 
\begin{subequations}
	\label{equ:evequsplitted}
	\begin{align}
	&\sum_{\nu'}\chi_{r,\text{A}}^{\nu\nu'\omega}V_{r,\text{A}}^{\nu'\omega}=\lambda_{r,\text{A}}^{\omega}V_{r,\text{A}}^{\nu\omega}\label{equ:evequsplittedanit}\\&\sum_{\nu'}\chi_{r,\text{S}}^{\nu\nu'\omega}V_{r,\text{S}}^{\nu'\omega}=\lambda_{r,\text{S}}^{\omega}V_{r,\text{S}}^{\nu\omega}.\label{equ:evequsplittedsymm}
	\end{align}
\end{subequations}
We will analyze these two eigenvalue problems separately in the following two subsections.

\subsection{Antisymmetric eigenvectors of \texorpdfstring{$\chi_r^{\nu\nu'\omega}$}{chi}}
\label{sec:eigantisymmetric}

The eigenvalue equation [Eq.~(\ref{equ:evequsplittedanit})] for the antisymmetric eigenvector explicitly reads as [see Eq.~(\ref{equ:chiAL})]
\begin{equation}
\label{equ:evequanti}
\sum_{\nu'}a_{0,r}^{\nu\omega}[\delta_{\nu\nu'}-\delta_{\nu(-\nu'-\omega)}]V_{r,\text{A}}^{\nu'\omega}=\lambda_{r,\text{A}}^{\omega}V_{r,\text{A}}^{\nu\omega.}
\end{equation}
Using the symmetry properties of $a_{0,r}^{\nu\omega}$ and $V_{r,\text{A}}^{\nu\omega}$ under the transformation $\nu\!\rightarrow\!-\nu-\omega$ one obtains the relation
\begin{equation}
\label{equ:evequanti1}
2a_{0,r}^{\nu\omega}V_{r,\text{A}}^{\nu\omega}=\lambda_r^{\omega}V_{r,\text{A}}^{\nu\omega},
\end{equation}
which corresponds to an eigenvalue equation of an already diagonal matrix. The normalized eigenvectors read as
\begin{equation}
\label{equ:eigenvectorantisymmetric}
V_{r,\text{A}}^{\nu\omega}=\frac{1}{\sqrt{2}}[\delta_{\nu\nu^*}-\delta_{\nu(-\nu^*-\omega)}].
\end{equation}
and the corresponding eigenvalues for the fixed frequency $\nu^*$ are given by [see Eq.~(\ref{equ:defchisplit1})]
\begin{equation}
\label{equ:eigenvalueantisymmetrix}
\lambda_{r,\text{A}}^{\omega}\equiv\lambda_{r,\text{A}}^{\nu^*\omega}=2a_{0,r}^{\nu^*\omega}=\frac{\mathcal{A}_0^r\beta[\nu^*(\nu^*+\omega)-A_r^2]}{[(\nu^*)^2+\frac{U^2}{4}][(\nu^*+\omega)^2+\frac{U^2}{4}]}.
\end{equation}
The eigenvalue $\lambda_{r,\text{A}}^{\nu^*\omega}$ clearly vanishes if $\nu^*(\nu^*\!+\omega)\!=A_r^2$, and the outer product of the eigenvector $V_{r,\text{A}}^{\nu\omega}$ is indeed identical to the shape of the corresponding localized divergences in $\Gamma_r^{\nu\nu'\omega}$ discussed in Sec.~\ref{subsec:divergchibar}.

\subsection{Symmetric eigenvectors of \texorpdfstring{$\chi_r^{\nu\nu'\omega}$}{chi}}
\label{sec:eigsymmetric}

The calculation of the symmetric eigenvectors $V_{r,\text{S}}^{\nu\omega}$ and the corresponding eigenvalues $\lambda_{r,\text{S}}^{\omega}$ is more difficult than in the antisymmetric case since the symmetric part of the susceptibility $\chi_{r,\text{S}}^{\nu\nu'\omega}$ exhibits nondiagonal terms. However, since its nondiagonal contribution corresponds to a matrix of only rank 2, we can follow a similar strategy as for the inversion of $\chi_r^{\nu\nu'\omega}$. The explicit eigenvalue equation [Eq.~(\ref{equ:evequsplittedsymm})] is given by
\begin{align}
\label{equ:evequsymm}
\sum_{\nu'} \{b_{0,r}^{\nu\omega}[\delta_{\nu\nu'}+\delta_{\nu(-\nu'-\omega)}]+\textstyle\sum\limits_{l=1}^2b_{l,r}^{\nu\omega}b_{l,r}^{\n'\omega}\}V_{r,\text{S}}^{\nu'\omega}=\lambda_{r,\text{S}}^{\omega}V_{r,\text{S}}^{\nu\omega}.
\end{align}
Using the symmetry of both $b_{0,r}^{\nu\omega}$ and $V_{r,\text{S}}^{\nu\omega}$ under the transformation $\nu\!\rightarrow\!-\nu-\omega$ one obtains the equivalent relation
\begin{subequations}
	\label{equ:evequsymm1}
	\begin{align}
	\sum_{l=1}^2b_{l,r}^{\nu\omega} P_{l,r}^{\omega} = & (-2b_{0,r}^{\nu\omega}+\lambda_{r,\text{S}}^{\omega}) V_{r,\text{S}}^{\nu\omega},\label{equ:evequsymm1a}\\
	P_{l,r}^{\omega} \equiv & \sum_{\nu'}b_{l,r}^{\nu'\omega}V_{r,\text{S}}^{\nu'\omega}.\label{equ:evequsymm1b} 
	\end{align}
\end{subequations}
Assuming that we know $P_{l,r}^{\nu\omega}$ and $\lambda_{r,\text{S}}^{\omega}$, the eigenvector $V_{r,\text{S}}^{\nu\omega}$ is in general given as
\begin{equation}
\label{equ:eigenvectorsymmetric1}
V_{r,\text{S}}^{\nu\omega}= {\mathbf P}\big[\frac{\sum_{l=1}^2 P_{l,r}^{\omega} b_{l,r}^{\nu\omega}}{\lambda_{r,\text{S}}^{\omega}-2b_{0,r}^{\nu\omega}}\big] + \delta(\lambda_{r,\text{S}}^{\omega}-2b_{0,r}^{\nu\omega})c_{r}^{\nu\omega},
\end{equation}
where $\mathbf P$ takes the principal value, i.e. excludes any point for which $\lambda_{r,\text{S}}^{\omega} = 2b_{0,r}^{\nu\omega}$, and $c_{r}^{\nu\omega}$ is a constant to be determined. 

Let us first consider the triplet channel $r\!=\!t$. In this case $b_{1,r}^{\nu\omega}\!=\!b_{2,r}^{\nu\omega}\!\equiv\!0$, which implies that $P_{l,r}^{\omega} = 0$. The symmetric eigenvectors of $\chi_t^{\nu\nu'\omega}$ are, hence, given by
\begin{equation}
\label{equ:evectortriplet}
V_{t,\text{S}}^{\nu\omega}=\frac{1}{\sqrt{2}}[\delta_{\nu\nu^*}+\delta_{\nu(-\nu^*-\omega)}],
\end{equation}
and the corresponding eigenvalue $\lambda_{t,\text{S}}^{\omega}$ reads
\begin{equation}
\label{equ:evaluetriplet}
\lambda_{t,\text{S}}^{\omega}\equiv\lambda_{t,\text{S}}^{\nu^*\omega}=2b_{0,t}^{\nu^*\omega}=\frac{-\beta\nu^*(\nu^*+\omega)}{2((\nu^*)^2+\frac{U^2}{4})((\nu^*+\omega)^2+\frac{U^2}{4})},
\end{equation}
which is always different from zero.

Now we consider Eq.~(\ref{equ:evequsymm1}) for the remaining cases $r\!=\!d,m,s$ where $b_{1,r}^{\nu\omega},b_{2,r}^{\nu\omega}\!\ne\!0$. Let us first analyze the case when $\lambda_{r,\text{S}}^{\omega} \neq 2b_{0,r}^{\nu\omega}$. This condition simplifies the eigenvector to the form
\begin{equation}
\label{equ:eigenvectorsymmetricp}
V_{r,\text{S}}^{\nu\omega}= \frac{\sum_{l=1}^2 P_{l,r}^{\omega} b_{l,r}^{\nu\omega}}{\lambda_{r,\text{S}}^{\omega}-2b_{0,r}^{\nu\omega}}.
\end{equation}
In order to determine the values of the quantities $P_{l,r}^{\omega}$ we substitute this expression into Eq.~(\ref{equ:evequsymm1b}), which yields the following homogeneous linear equation for $P_{l,r}^{\omega}$:
\begin{subequations}
\label{equ:eigenvectorsymmetric2}
	\begin{align}
	\sum_{l=1}^2 L_{kl}^{r,\omega}&(\lambda_{r,\text{S}}^{\omega}) P_{l,r}^{\omega}=  0,\label{equ:eigenvectorsymmetric2a}\\L_{kl}^{r,\omega}(\lambda_{r,\text{S}}^{\omega}) = \delta_{kl}&+\sum_{\nu}\frac{b_{k,r}^{\nu\omega}b_{l,r}^{\nu\omega}}{2b_{0,r}^{\nu\omega}-\lambda_{r,\text{S}}^{\omega}}.\label{equ:eigenvectorsymmetric2b}
	\end{align}
\end{subequations}
Clearly, Eq.~(\ref{equ:eigenvectorsymmetric2a}) has only a nontrivial solution if $L_{kl}^{r,\omega}(\lambda_{r,\text{S}}^{\omega})$ becomes singular. Formally, this corresponds to the condition Det$[L_{kl}^{r,\omega}(\lambda_{r,\text{S}}^{\omega})]\!=\!0$ which represents a transcendental equation for $\lambda_{r,\text{S}}^{\omega}$. After the value of $\lambda_{r,\text{S}}^{\omega}$ has been determined by means of this equation, Eqs.~(\ref{equ:eigenvectorsymmetric2}) can be solved for $P_{l,r}^{\omega}$ (or, more precisely, for the ratio  $P_{1,r}^{\omega}/ P_{2,r}^{\omega}$) which yields the final expression for $V_{r,\text{S}}^{\nu\omega}$ in Eq.~(\ref{equ:eigenvectorsymmetricp}).

The matrix $L_{kl}^{r,\omega}(\lambda_{r,\text{S}}^{\omega})$ is similar to $M_{kl}^{r,\omega}$ in Eq.~(\ref{equ:chibarfinal}) and can be in principle calculated analytically. However, for $\lambda_{r,\text{S}}^{\omega}\!\ne\!0$ the structure of the poles of the summand in the Matsubara frequency sum in Eq.~(\ref{equ:eigenvectorsymmetric2}) is considerably more complicated compared to the corresponding expression for $M_{kl}^{r,\omega}$ and we, hence, refer the reader to Appendix~\ref{app:expliciteigen} for the final (rather lengthy) results.

The most interesting case is, however, when $\lambda_{r,\text{S}}^\omega\!=\!0$ which signals a {\em global} divergence of the corresponding vertex $\Gamma_r^{\nu\nu'\omega}$. In this situation, the matrix $L_{kl}^{r,\omega}(\lambda_{r,\text{S}}^{\omega}\!=\!0)$ becomes equivalent to $M_{kl}^{r,\omega}$ whose determinant vanishes when the condition in Eq.~(\ref{equ:gammaglobdiv}) is fulfilled. In fact, all matrix elements $L_{kl}^{r,\omega}(\lambda_{r,\text{S}}^{\omega}\!=\!0)=M_{kl}^{r,\omega}$ become $0$ except for $L_{11}^{r,\omega}(\lambda_{r,\text{S}}^\omega\!=\!0)$ and, hence, $P_{1,r}^{\omega}\!=\!0$. This is indeed consistent with the requirement that $V_{r,\text{S}}^{\nu\omega}$ should be normalizable which would not be the case for the term $b_{1,r}^{\nu\omega}/b_{0,r}^{\nu\omega}$ which appears for $\lambda_{r,\text{S}}^{\omega}\!=\!0$ in Eq.~(\ref{equ:eigenvectorsymmetricp}). The eigenvector to a vanishing eigenvalue is therefore $\propto b_{2,r}^{\nu\omega}/b_{0,r}^{\nu\omega}$ and explicitly reads as
\begin{equation}
\label{equ:eigenvector0}
V_{r,\text{S}}^{\nu\omega}\lvert_{\lambda_{r,\text{S}}^{\omega}=0}=\frac{N}{\nu(\nu+\omega)-B_r^2},
\end{equation}
where $N$ can be easily obtained from the normalization condition $\sum_\nu (V_{r,\text{S}}^{\nu\omega})^2\!=\!1$. The outer product of this eigenvector indeed reproduces the globally divergent component in $\Gamma_r^{\nu\nu'\omega}$ in the second line of Eq.~(\ref{equ:gammafinal}).

Let us finally consider the case when $\lambda_{r,\text{S}}^{\omega} = 2b_{0,r}^{\nu^*\omega}$, for some fixed frequency $\nu^*$. This condition reduces Eq.~(\ref{equ:evequsymm1a}), evaluated at $\nu^*$, to the constraint
\begin{equation}
\label{equ:evequsymmconst}
P_{2,r}^{\omega} = -\frac{b_{1,r}^{\nu^*\omega}}{b_{2,r}^{\nu^*\omega}}P_{1,r}^{\omega}.
\end{equation}
Substituting Eqs.~(\ref{equ:eigenvectorsymmetric1}) and (\ref{equ:evequsymmconst}) into Eq.~(\ref{equ:evequsymm1b}) gives now {\em two} (nontrivial) linear equations ($l=1,2$) but only {\em one} free parameter $c_{r}^{\nu^*\omega}/P_{1,r}^{\omega}$. This implies that there are in general no solutions fulfilling $\lambda_{r,\text{S}}^{\omega} = 2b_{0,r}^{\nu^*\omega}$ for a fixed frequency $\nu^*$ and interaction strength $U$. Instead, any valid solution of this type can be reached by intersecting the condition in Eq.~(\ref{equ:evequsymmconst}) with the {\em closure} of the solutions with $\lambda_{r,\text{S}}^{\omega} \neq 2b_{0,r}^{\nu^*\omega}$. In particular, for $\lambda_{r,\text{S}}^{\omega} = 0$ there are no solutions for the condition (\ref{equ:gammaglobdiv}) in the neighborhood of $b_{0,r}^{\nu^*\omega} = 0$, for any finite frequency $\nu^*$. This implies that $\chi_{r,\text{S}}^{\nu\nu'\omega}$ has no zero at the points $b_{0,r}^{\nu^*\omega}=0$ and, hence, any divergence in $\Gamma_r^{\nu\nu'\omega}$ at these points must cancel (c.f. Sec.~\ref{subsec:divergchibar}). 

\section{Approximate self-energy functional \texorpdfstring{$\Sigma[G]$}{Sigma[G]}}
\label{sec:LW}

In DMFT the lattice problem is mapped onto an Anderson impurity model through the noninteracting one-particle bath Green's function $G_0$. The impurity solver can formally be seen as functional $G[G_0]$ that for each $G_0$ returns the corresponding fully interacting local Green's function $G$. This functional is in general {\em not} injective\cite{Kozik2015,Gunnarsson2017}, i.e., there are several different noninteracting bath Green's functions $G_0$ that can produce the same interacting Green's function $G$. This implies a multivaluedness\cite{Kozik2015,Rossi2015,Stan2015,Schafer2016,Tarantino2017} of the inverse functional $G_0[G]$ and, via the Dyson equation, of the self-energy functional $\Sigma[G]$. Fortunately, there can be at most one of these $G_0$'s, which we will call the physical $G_0^{\text{phys}}$, that corresponds to a noninteracting impurity problem.\cite{Potthoff2003,Potthoff2006} 

In Refs.~\onlinecite{Kozik2015} and \onlinecite{Gunnarsson2017} it has been numerically shown for the AL that at specific values of $U$, $G_0^{\text{phys}}$ and another $G_0$ become identical (cross) and that such a crossing implies the divergence of the irreducible density vertex $\Gamma_d$ (see SM in Ref.~\onlinecite{Gunnarsson2017}). 

Analytically such a scenario has been first demonstrated for the simple cases of the one-point model\cite{Rossi2015,Stan2015} and  disordered systems\cite{Janis2014,Schafer2016} such as the BM in infinite dimensions, where explicit expressions for the functional $G[G_0]$ and the irreducible vertices $\Gamma_r$ are available. 

On the contrary, for the AL of the Hubbard model no analytical expression for the exact functional $G[G_0]$ is known. In Ref.~\onlinecite{Gunnarsson2017} some of the present authors used a numerically exact quantum Monte Carlo solver to obtain the different $G_0$'s which yield the physical $G$ of Eq.~(\ref{equ:1pgf}). In this paper we will follow a complementary path. Instead of adopting a numerically exact solver we will use the Dyson equation
\begin{equation}
\label{equ:dyson}
G[G_0] = (G_0^{-1}\!-\!\Sigma[G_0])^{-1},
\end{equation}
and approximate the self-energy functional $\Sigma[G_0]$ by the general form of the Iterated Perturbation Theory\cite{Georges1996,Held2007} (IPT) expression (which is the same as for the full Hubbard model)
\begin{align}
\label{equ:approxfunct}
&\Sigma^{a}_\sigma[G_0](\nu) = -\frac{U}{2}  + \frac{U}{\beta}\sum_{\nu_1}G_{0,(-\sigma)}(\nu_1)\nonumber\\
& -\frac{U^2}{\beta^2} \sum_{\nu_1\omega} \!G_{0,(-\sigma)}(\nu_1)G_{0,(-\sigma)}(\nu_1+\omega)G_{0,\sigma}(\nu+\omega).
\end{align}
$\Sigma^{a}_\sigma[G_0](\nu)$ yields the exact self-energy of the Hubbard atom when it is evaluated with $G_0^{\text{phys}}(\nu)\!=\!1/i\nu$. Within DMFT, it captures strong coupling phenomena such as the Mott metal-to-insulator transition.\cite{Georges1996} The analytical form allows us to investigate which ingredients cause the emergence of different types of divergences in $\Gamma_r$. In particular, we will stress the differences between disordered and fully interacting systems such as the Hubbard atom w.r.t. these singularities. In the following, we will restrict ourselves, for simplicity, to the case of a spin independent $G_{0,\sigma}(\nu)\!=\!G_0(\nu)$ without any anomalous contributions, as in the numerical calculations of Ref.~\onlinecite{Gunnarsson2017}. This limits the following analysis to the density channel.\footnote{The calculation of a response function $\chi_r$ by means of the functional derivative $\delta G/\delta G_0$ [see Eqs.~(\ref{equ:funcder})] requires the introduction of a symmetry-breaking field in channel $r$.} 

\subsection{Unphysical \texorpdfstring{$G_0$}{G0} solutions for the IPT functional}
\label{sec:unphys}

Substituting Eq.~(\ref{equ:approxfunct}) into Eq.~(\ref{equ:dyson}) gives $N$ coupled fourth-order equations in $G_0(\nu)$ where $N$ is the number of Matsubara frequencies which we consider. Hence, one has to expect $4^N$ solutions for $G_0(\nu)$ for a given $G(\nu)$. For the numerical calculation we have fixed $G(\nu)$ to the physical Green's function [Eq.~(\ref{equ:1pgf})], $N\!=\!8$, $0\!<\!U\!<\!2.6$, $\beta\!=\!2$, and set $G_0(\nu)$ for all frequencies $\lvert\nu\rvert\!>\!15\pi/\beta$ to its physical value $1/i\nu$. In order to obtain different solutions of Eq.~(\ref{equ:dyson}) we start with an initial guess for $G_0(\nu)$. By means of a Metropolis search we find an improved guess around which we repeatedly linearize Eq.~(\ref{equ:dyson}) [together with Eq.~(\ref{equ:approxfunct})] until a $G_0(\nu)$ is found which reproduces the physical $G(\nu)$ up to a given accuracy. This way, it is possible to identify different unphysical $G_0(\nu)$ with a $1/i\nu$ asymptotic high-frequency behavior.

\begin{figure}[t!]
	\centering
	\includegraphics[width=\columnwidth]{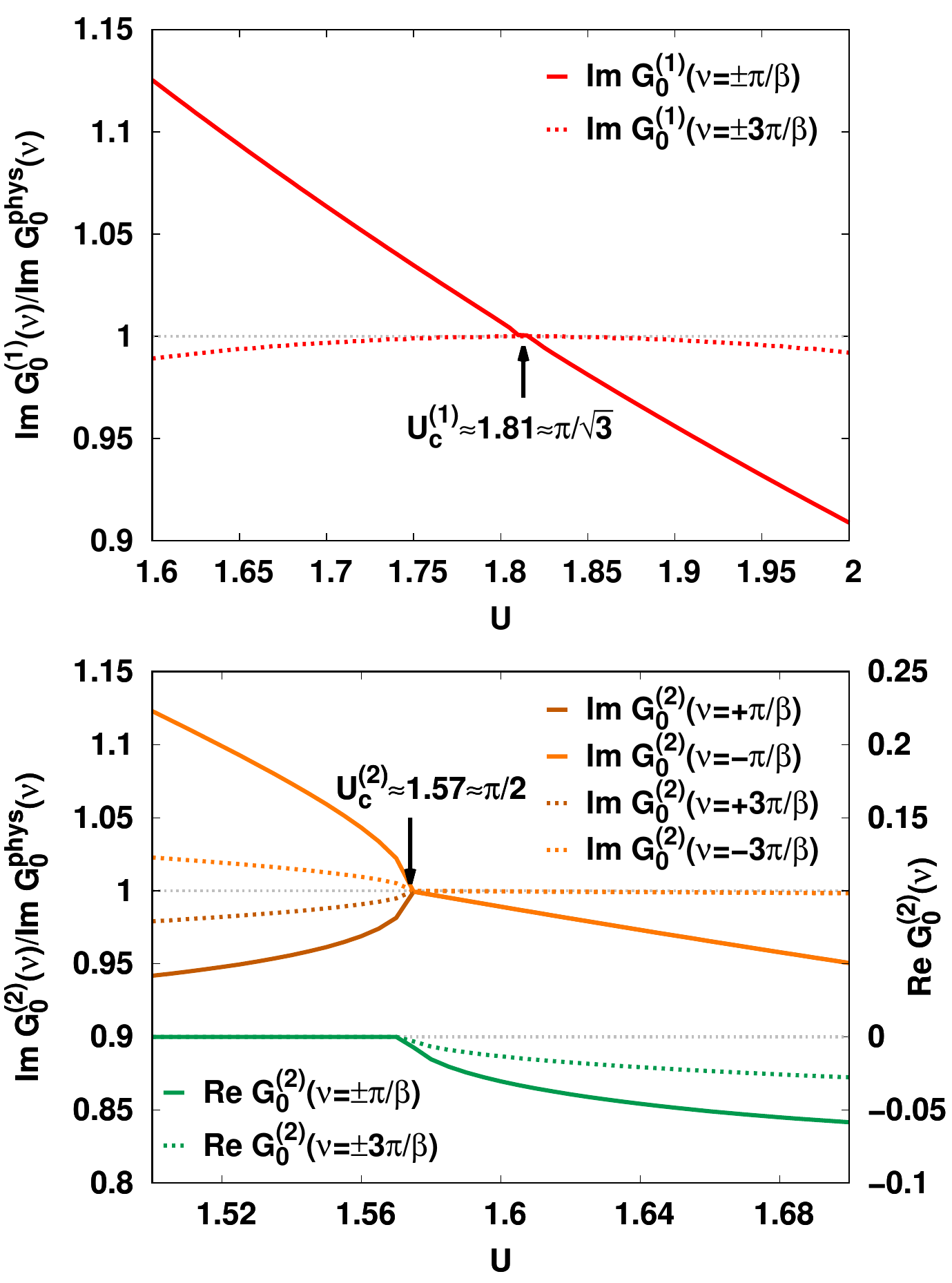}
	\caption{First two crossings of the physical $G_0^{\text{phys}}(\nu)$ with two unphysical ones [$G_0^{(1)}(\nu)$ and $G_0^{(2)}(\nu)$] as a function of $U$ at the first two Matsubara frequencies ($\nu_1\!=\!\pm\pi/\beta$ and $\nu_2\!=\!\pm3\pi/\beta$). The imaginary part is normalized by $\text{Im} G_0^{\text{phys}}(\nu)$. Note that for the red solution [$G_0^{(1)}(\nu)$] the relation $[G_0^{(1)}]^*(\nu)\!=\!G_0^{(1)}(-\nu)\!=\!-G_0^{(1)}(\nu)$ holds and, hence, the results for positive and negative Matsubara frequencies are the same. For the orange solution $G_0^{(2)}(\nu)$ this is only true after the crossing with the physical $G_0^{\text{phys}}(\nu)$ where it, however, acquires a real part. $\text{Im}G_0^{(1)}(\nu=\pm3\pi/\beta)/\text{Im}G_0^{\text{phys}}(\nu=\pm3\pi/\beta)$ in the upper panel has been rescaled by a factor of $10$ for a better visibility. $\beta\!=\!2$.}
	\label{fig:Crossing_G0_orange}
\end{figure}

For the $U$ range considered in our numerical calculationsm we find two unphysical $G_0^{(1)}(\nu)$ and $G_0^{(2)}(\nu)$ which become identical to (cross) the physical $G_0^{\text{phys}}(\nu)$ at $U_c^{(1)} \!=\!\pi/\sqrt{3}$ and $U_c^{(2)} \!=\! \pi/2$, respectively. The unphysical nature of $G_0^{(1)}(\nu)$ and $G_0^{(2)}(\nu)$ is reflected in an {\em increase} of double occupancy with {\em increasing} $U$ as shown in Appendix~\ref{app:double}. The direct comparison of the physical with the two unphysical $G_0(\nu)$'s in Fig.~\ref{fig:Crossing_G0_orange} is similar to the corresponding results of Ref.~\onlinecite{Gunnarsson2017} (where an exact Monte Carlo solver has been used for the functional $G[G_0]$), except for the value of $U_c^{(2)}$. In the upper panel we show the imaginary part of $G_0^{(1)}(\nu)$ normalized by the imaginary part of the physical $G_0^{\text{phys}}(\nu)$ for the first two Matsubara frequencies as a function of $U$. While for $\nu\!=\!\pm\pi/\beta$ the unphysical $G_0^{(1)}(\nu)$ crosses the physical one linearly in $U\!-\!U_c^{(1)}$, for the higher Matsubara frequencies the crossing occurs quadratically. Let us note that $G_0^{(1)}(\nu)$ is purely imaginary and fulfills the standard relation for complex conjugation which renders the result for the positive and negative frequencies equivalent. In the lower panel of Fig.~\ref{fig:Crossing_G0_orange}, $G_0^{(2)}(\nu)$ is shown as a function of $U$ for the two lowest Matsubara frequencies. The imaginary part is again normalized by $\text{Im}G_0^{\text{phys}}(\nu)$ while the real part is plotted in absolute values. $G_0^{(2)}(\nu)$ crosses the physical $G_0^{\text{phys}}(\nu)$ at $U_c^{(2)}\!~\approx\!\pi/2$. Before this crossing the relation $G_0^*(\nu)\!=\!G_0(-\nu)$ is violated for $G_0^{(2)}(\nu)$ and, hence, we obtain different results for positive and negative Matsubara frequencies. For $U\!>\!U_c^{(2)}$, on the other hand, this relation is fulfilled but $G_0^{(2)}(\nu)$ acquires a finite real part. Our numerical results indicate that for $G_0^{(2)}(\nu)$ the crossing with $G_0^{\text{phys}}(\nu)$ is proportional to $\sqrt{U-U_c}$ for {\em all} frequencies, within the imposed numerical accuracy.

Let us stress that our findings coincide (at least qualitatively) with the numerically exact results of Ref.~\onlinecite{Gunnarsson2017}. This demonstrates the applicability of our approximate IPT functional to analyze the multivaluedness of $G_0[G]$ and its connection to the divergences of $\Gamma_r$ in the Hubbard atom.

\subsection{IPT susceptibilities}
\label{sec:1pisusc}

The generalized susceptibility in the density channel at the bosonic Matsubara frequency $\omega\!=\!0$ ($\chi_d^{\nu\nu'(\omega=0)}$) can be calculated from $G[G_0]$ in Eq.~(\ref{equ:dyson}) as\footnote{Note that this result yields the same $\Gamma_d$ as it would be obtained from the functional derivative $\delta\Sigma / \delta G$ (we neglect frequency and spin arguments for the sake of simplicity). In fact, inverting the functional $G[G_0]$ and using the Dyson Eq.~(\ref{equ:dyson}), we have $\Sigma[G]=G_0^{-1}[G]-G^{-1}$. Taking the derivative with respect to $G$ yields $\delta\Sigma / \delta G = G^{-2}+\delta G_0^{-1} / \delta G = G^{-2}+ [\delta G / \delta G_0^{-1}]^{-1} = \beta(-\chi_0^{-1}+\chi^{-1}) = \Gamma/\beta$ as defined in Eq.~(\ref{equ:BSinv}).} $\chi_d^{\nu\nu'(\omega=0)}\!=\beta\frac{\delta G_\uparrow(\nu)}{\delta G_{0,\uparrow}^{-1}(\nu')}\!+\!\beta\frac{\delta G_\uparrow(\nu)}{\delta G_{0,\downarrow}^{-1}(\nu')}$. Using the Dyson equation~(\ref{equ:dyson}) and the IPT self-energy functional in Eq.~(\ref{equ:approxfunct}), these derivatives can be straightforwardly performed. Since we are interested in the divergences of the physical branch, we evaluate the resulting expressions at the spin-independent physical $G_0^{\text{phys}}(\nu)$ and $G(\nu)$ which yields
\begin{subequations}
	\label{equ:funcder}
	\begin{align}
	\label{equ:funcder1}
	\frac{\delta G_\uparrow(\nu)}{\delta G_{0,\uparrow}^{-1}(\nu')}=&-G^2(\nu)[1+\frac{U^2}{4}G_0^{2}(\nu)]\delta_{\nu\nu'},\\
	\frac{\delta G_\uparrow(\nu)}{\delta G_{0,\downarrow}^{-1}(\nu')}=&-\frac{U^2}{4}G_0^{2}(\nu)G^2(\nu)[\delta_{\nu\nu'}-\delta_{\nu(-\nu')}]\nonumber\\&-\frac{U}{\beta}G^{2}(\nu)G^2_0(\nu'),\label{equ:funcder2}
	\end{align}
\end{subequations}
where we have used that at half-filling $G_0(-\nu)^2\!=\!G_0(\nu)^2$. The nondiagonal term on the right-hand side of Eq.~(\ref{equ:funcder2}) arises directly from the (bare) Hartree contribution in $\Sigma_{\sigma}^a[G_0]$. $\chi_d^{\nu\nu'(\omega=0)}$ now becomes
\begin{align}
\label{equ:approxsusc}
\chi_d^{\nu\nu'(\omega=0)}=&-\beta\frac{G^2(\nu)}{2}\left[1+\frac{3U^2}{4}G_0^2(\nu)\right][\delta_{\nu\nu'}-\delta_{\nu(-\nu')}]\nonumber\\&-\beta\frac{G^2(\nu)}{2}\left[1+\frac{U^2}{4}G_0^2(\nu)\right][\delta_{\nu\nu'}+\delta_{\nu(-\nu')}]
\nonumber\\&-U G^2(\nu)G_0^2(\nu').
\end{align}
One can clearly see that this susceptibility has the form of Eq.~(\ref{equ:chiAL}) with the only difference that it consists of only one nondiagonal term. Furthermore, we see that the generalized susceptibility produced by IPT is not symmetric with respect to $\nu\!\leftrightarrow\!\nu'$, which means it violates time-reversal symmetry. We can, nevertheless, use Eq.~(\ref{equ:chibarfinal}) to calculate the inverse of $\chi_d^{\nu\nu'(\omega=0)}$, as given in Eq.~(\ref{equ:approxsusc}), which yields
\begin{align}
\label{equ:approxsuscinv}
\overline{\chi}_d&^{\nu\nu'(\omega=0)}=-\frac{1}{2\beta G^2(\nu)}\frac{1}{1-\frac{3U^2}{4\nu^2}}[\delta_{\nu\nu'}-\delta_{\nu(-\nu')}]\nonumber\\&-\frac{1}{2\beta G^2(\nu)}\frac{1}{1-\frac{U^2}{4\nu^2}}[\delta_{\nu\nu'}+\delta_{\nu(-\nu')}]\nonumber\\&+\frac{U}{\beta^2}\frac{1}{1-\tan(\frac{\beta U}{4})}\left[\frac{G_0(\nu')}{G(\nu)}\right]^2\frac{1}{1-\frac{U^2}{4\nu^2}}\frac{1}{1-\frac{U^2}{4\nu'^2}}.
\end{align}

Let us now analyze the possible divergences in Eq.~(\ref{equ:approxsuscinv}): 

(i) In the first line of Eq.~(\ref{equ:approxsuscinv}) we encounter a {\em local} divergence when $\nu\!=\!\nu^*\!=\!\frac{U}{2}\sqrt{3}$ which is equivalent to the one found by the exact calculation for the atomic limit in Eq.~(\ref{equ:reddiv}). Hence, the approximate self-energy functional of IPT indeed reproduces correctly the localized divergences in the AL at $\omega\!=\!0$. Moreover, consistent with the proof in Ref.~\onlinecite{Gunnarsson2017}, the divergence occurs only for the frequencies for which the crossing of the physical and unphysical $G_0$'s is of lowest order, in this case linear. Just as for the exact susceptibility, the factor $\sqrt{3}$ originates from adding the two spin combinations $\uparrow\uparrow$ and $\uparrow\downarrow$ in Eqs.~(\ref{equ:funcder}). Let us remark that this result can be even obtained from a  perturbative low-order expansion of $\chi_d^{\nu\nu'\omega}$ in $U$ as long as the SU(2) symmetry of the system is correctly taken into account.

(ii) In lines two and three of Eq.~(\ref{equ:approxsuscinv}) we would expect a divergence at $\nu\!=\!\frac{U}{2}$, analogous to an apparent singularity at $B_r$ for the exact solution in Eq.~(\ref{equ:red2div}) (which actually becomes $B_r\!\rightarrow\!\frac{U}{2}$ in the perturbative limit $U\rightarrow 0$). As discussed already  in Sec.~\ref{sec:gamma}, these divergences cancel. This is also consistent with the fact that no unphysical $G_0$ crosses the physical $G_0$ at $U\!=\!\pi$ for $\beta\!=\!2$. 

(iii) If $\tan(\frac{\beta U}{4})\!=\!1$ in the denominator in the last line of Eq.~(\ref{equ:approxsuscinv}) we will get a {\em global} divergence of $\overline{\chi}_d$. This expression is not the same as the exact result in Eq.~(\ref{equ:gammaglobdiv}), but instead consistent with the observed crossing of $G_0$'s in the lower panel of Fig.~\ref{fig:Crossing_G0_orange} obtained from IPT. The global nature of the divergence is also consistent with the fact that the corresponding crossing seems to happen for {\em all} Matsubara frequencies in the same way ($\propto\!\!\sqrt{U-U_c}$). Following Eqs.~(\ref{equ:approxsusc}) and (\ref{equ:approxsuscinv}), we can trace the origin of the global divergence back to the Hartree term in Eq.~(\ref{equ:approxfunct}). In fact, global singularities can only arise from nondiagonal contributions to the generalized susceptibility $\chi_r^{\nu\nu'\omega}$ which originate from the functional derivative of the Hartree term in the third line of Eq.~(\ref{equ:approxsusc}). In this respect, a global vertex divergence of $\Gamma_r^{\nu\nu'\omega}$ marks an important difference between fully interacting and disordered systems, where due to the absence of an interaction among the particles itself no global divergence can be found\cite{Janis2014,Schafer2016}. Let us point out that this is, however, only true when the local potential is considered to be random but fixed, i.e., a static (quenched) disorder. On the contrary, for systems with annealed disorder, such as the Falicov-Kimball model, the value of the local potential depends on the state of the system which implies that $\chi_r$ has nondiagonal terms\cite{Schafer2016}.

\section{Interpretation of the vertex divergences} 
\label{sec:interpretation}

Let us briefly recall possible interpretations of the divergences of $\Gamma_r^{\nu\nu'\omega}$, the corresponding vanishing of an eigenvalue $\lambda_r^{\omega}$ of the generalized susceptibility $\chi_r^{\nu\nu'\omega}$ and the multivaluedness of the functional $\Sigma[G]$ (or $G_0[G]$). Formally, the appearance of the first singularity in $\Gamma_r^{\nu\nu'\omega}$ marks the point where any finite-order perturbation expansion for the two-particle vertex functions leads to, even qualitatively, wrong results (see, e.g., Fig.~1 in Ref.\onlinecite{Schafer2013}). At the same time, so-called ``bold'' diagrammatic methods (e.g., diagrammatic Monte Carlo\cite{Prokofev1998}) which self-consistently evaluate $\Sigma[G]$ (or $G_0[G]$) can converge to the unphysical branch of this functional which might restrict such approaches to the weak-coupling regime (for possible solutions of this problem, see Ref.~\onlinecite{Rossi2016}). From a physical perspective, the emergence of singularities in $\Gamma_r^{\nu\nu'\omega}$ has been associated\cite{Schafer2013} with the formation of Hubbard subbands in the spectral function, the emergence of ``kinks'' in the electronic self-energy\cite{Byczuk2007}, or a change in the relaxation mechanisms after a quench of the Hubbard interaction\cite{Eckstein2009}. All these attempts to explain the singularities rely, however, more or less on the observation that the interaction values at which these phenomena occur coincide with the appearance of divergences, rather than establishing a causal relation between these findings. In Ref.~\onlinecite{Gunnarsson2017}, on the other hand, the emergence of negative eigenvalues in the charge channel in the DMFT solution of the $2d$ half-filled Hubbard model was shown to suppress the local charge susceptibility (see also Ref.~\onlinecite{Gunnarsson2016}).

While no comprehensive physical interpretation of the individual eigenvalues of the generalized susceptibilities, and the corresponding vertex divergences, has been found so far, the question remains nevertheless very important. Indeed, if the eigenvalues have a physical interpretation, it becomes important to use approximations that preserve their character, i.e., (implicitly) allow for vertex divergences. On the other hand, if only a few linear combinations of the eigenvalues carry physical information, such as their total sum, then a wider set of approximations which suppress singularities of the vertex might become applicable.

\section{Relation to \texorpdfstring{$G(\nu)$}{G(nu)} in the Binary Mixture}
\label{sec:1pGF}

In this section, we want to comment briefly on the connection between the divergences of $\Gamma_d^{\nu\nu'\omega}$ and the minimum of the single-particle Green's function $G(\nu)$ in the disordered binary mixture model. It has been demonstrated\cite{Schafer2016} that in the BM a localized vertex divergence in $\Gamma_d^{\nu\nu'(\omega=0)}$ occurs at a frequency $\nu^*_{\text{BM}}\!=\!\nu_{\text{BM}}^{\text{min}}\!=\!U/2$ where the corresponding single-particle Matsubara Green's function $G_{\text{BM}}(\nu)$ exhibits a minimum. To prove this connection in a more general way, we note that the self-energy functional of the BM model is local in Matsubara frequencies, i.e., $\Sigma_{\text{BM}}(\nu)\!=\!\Sigma_{\text{BM}}[G_{\text{BM}}(\nu)]$. Consequently, the functional derivative $\delta \Sigma_{\text{BM}}[G_{\text{BM}}(\nu)] / \delta G_{\text{BM}}(\nu)\!=\!\frac{1}{\beta}\Gamma_{d,\text{BM}}^{\nu\nu(\omega=0)}$ corresponds to a normal derivative. Considering the Dyson equation for the BM in its atomic limit
\begin{equation}
\label{equ:1pgffunc}
G_{\text{BM}}(\nu)=\frac{1}{i\nu+\mu-\Sigma_{\text{BM}}[G_{\text{BM}}(\nu)]},
\end{equation}
and differentiating this relation with respect to $\nu$, we obtain
\begin{equation}
\label{equ:1pgfuncdiffBM}
\frac{\partial G_{\text{BM}}}{\partial \nu}(\nu)=-G_{\text{BM}}^2(\nu)\Bigl[i-\frac{1}{\beta}\Gamma_{d,\text{BM}}^{\nu\nu(\omega=0)}\frac{\partial G_{\text{BM}}}{d\nu}(\nu)\Bigr].
\end{equation}
Solving this equation for $\partial G_{\text{BM}}/\partial\nu$ yields
\begin{equation}
\label{equ:1pgfuncdiffBMrew}
\frac{\partial G_{\text{BM}}}{\partial \nu}(\nu)=-\frac{i}{G_{\text{BM}}^{-2}(\nu)-\frac{1}{\beta}\Gamma_{d,\text{BM}}^{\nu\nu(\omega=0)}}.
\end{equation}
Obviously, for a divergence of $\Gamma_{d,\text{BM}}^{\nu\nu(\omega=0)}$ at $\nu\!=\!\nu^*_{\text{BM}}$ the right hand side of this equation vanishes which implies that $\partial G_{\text{BM}}/d\nu(\nu^*_{\text{BM}})\!=\!0$. This proves that the single-particle Green's function of the BM exhibits an extremal point, in this case a minimum, exactly at the frequency $\nu_{\text{BM}}^{\text{min}}\!=\!\nu_{\text{BM}}^*$ where the corresponding density vertex diverges. This holds also in the case of a finite dispersion when the system is treated in the framework of CPA (see Appendix~\ref{app:inflection} and Ref.~\onlinecite{Schafer2016}).

Finally, on a more speculative note, let us point out that the second derivative of $G_{\text{BM}}(\nu)=G(\nu)$ [Eq.~(\ref{equ:1pgf})] with respect to $\nu$, in the atomic limit, gives
\begin{equation}
\label{equ:1pgf2nder}
\frac{\partial^2G_{\text{BM}}}{\partial\nu^2}(\nu)=-i\nu\frac{\nu^2-\frac{3U^2}{4}}{(\nu^2+\frac{U^2}{4})^3}.
\end{equation}
It vanishes precisely at $\nu^{\text{infl}}\!=\!\nu^*\!=\!\sqrt{3}U/2$, which coincides with the energy scale of $\chi_d^{\nu\nu'(\omega=0)}$ of the Hubbard atom in Eq.~(\ref{equ:reddiv}). An interesting future research question is how well this ``accidental'' relation between the inflection point of $G_{\text{BM}}(\nu)$ and $\chi_d^{\nu\nu'(\omega=0)}$ holds in the case of finite band width\cite{Chalupa2018}. 

\section{Conclusions and Outlook}
\label{sec:conclusions}

In this work, we have presented analytical expressions for the irreducible vertex $\Gamma_r^{\nu\nu'\omega}$, the fully irreducible vertex $\Lambda^{\nu\nu'\omega}$, and the eigenvalues and eigenvectors of the generalized susceptibilities $\chi_r^{\nu\nu'\omega}$, of the half-filled Hubbard atom. In spite of its simple Hamiltonian, the corresponding vertices exhibit a complex frequency structure and even capture the low-frequency singularities which have been discovered already for the DMFT solution of the Hubbard model. In order to gain further insight into the origin of these divergences, we have classified {\em all} eigenvalues and eigenvectors of the generalized susceptibilities in all parquet channels ($r$) and for all bosonic Matsubara frequencies $\omega$. The low-frequency divergences of $\Gamma_r^{\nu\nu'\omega}$ occur when these eigenvalues pass through zero while their frequency dependence is determined by the outer product of the corresponding eigenvectors. Specifically, we have identified vanishing eigenvalues associated with antisymmetric eigenvectors which correspond to {\em localized} singularities of $\Gamma_r^{\nu\nu'\omega}$ at $\nu(\nu+\omega)\!=\!A_r^2$ and, hence, set an energy scale at which a perturbative treatment breaks down\cite{Schafer2016}. Vanishing eigenvalues with symmetric eigenvectors correspond to {\em global} divergences of the irreducible vertex. Nevertheless, a definite physical interpretation of both types of singularities of $\Gamma_r^{\nu\nu'\omega}$ and the corresponding crossing of eigenvalues of $\chi_r^{\nu\nu'\omega}$ through $0$ has not been achieved so far. However, we have shown that the divergences of $\Gamma_r^{\nu\nu'\omega}$ in the Hubbard atom can be modeled {\em qualitatively} by using the self-energy functional of iterated perturbation theory. In this way, we could identify the (bare) Hartree term to be essential for the emergence of global divergences. This also marks an important difference to systems with quenched disorder such as the BM where the interaction between the particles and, hence, global divergences in $\Gamma_r^{\nu\nu'\omega}$, are absent.

We hope that our analytical results can guide the development of new approximation schemes for the vertex functions and the BS equations of more complex correlated systems, such as the Hubbard model, and possibly allow for a more comprehensive understanding of strong-coupling phenomena in many electron models and correlated materials. For instance, one might use the spectral representation of the irreducible vertex, truncated at a finite (low) number of eigenvalues, to solve the BS equation for obtaining the generalized susceptibilities and corresponding response functions in a semi-analytical way, following the procedure shown in Sec.~\ref{sec:gamma}.

Our analytical derivations have been simplified by the particle-hole symmetry of the half-filled Hubbard atom. In realistic material calculations, it is, however, rare that particle-hole symmetry is fulfilled, which makes a study of the Hubbard atom away from half-filling an interesting future research direction. In addition, if the system undergoes a spontaneous SU(2) symmetry breaking, the BS equations cannot be decoupled into spin-singlet and spin-triplet (i.e., density and spin as well as particle-particle singlet and particle-particle triplet) components. Whether such a symmetry breaking can mitigate the emergence of divergences in the irreducible vertex is an open question which also requires further investigation. The analytical formulas for the two-particle Green's function in Ref.~\onlinecite{Pairault2000} can serve as an excellent starting point for both these studies. 

\paragraph*{Acknowledgements}
We thank P. Chalupa, K. Held, F. Kugler, T. Ribic, A. N. Rubtsov, D. Springer, C. Taranto and A. Toschi for useful discussions. We acknowledge financial support from the Russian Science Foundation through Grant No.~16-42-01057 (GR).

\appendix

\section{Symmetry decomposition}
\label{app:symmetries}
\begin{table}
	\caption{Symmetry relations of the generalized susceptibilities $\chi_{\sigma\sigma'}^{\nu\nu'\omega}$. The particle-hole and particle-particle notations are only indicated when the results differ. Note that the frequency shift needed to switch between the particle-hole and the particle-particle notation is defined as $\omega \rightarrow -\omega -\nu -\nu'$.}\label{tab:symmetry}
	{\renewcommand{\arraystretch}{2.0}
		\begin{tabular}{ c l }
			\hline
			\bf{Symmetry} & \bf{Relation} \\
			\hline
			\parbox[c][1.0cm]{3cm}{Complex Conjugation} & 
			{$\chi_{\sigma\sigma'}^{\nu\nu'\omega} = (\chi_{\sigma'\sigma}^{(-\nu')(-\nu)(-\omega)})^*$} \\
			\parbox[c][1.0cm]{3cm}{Swapping (ph)}& 
			{$\chi_{ph,\sigma\sigma'}^{\nu\nu'\omega} = \chi_{ph,\sigma'\sigma}^{(\nu'+\omega)(\nu+\omega)(-\omega)}$} \\ 
			\parbox[c][1.0cm]{3cm}{Swapping (pp)}& 
			{$\chi_{pp,\sigma\sigma'}^{\nu\nu'\omega} = \chi_{pp,\sigma'\sigma}^{(-\nu-\omega)(-\nu'-\omega)\omega}$} \\
			\parbox[c][1.0cm]{3cm}{Spin-SU(2)}& 
			{$\chi_{\sigma\sigma'}^{\nu\nu'\omega} = \chi_{\sigma'\sigma}^{\nu\nu'\omega} = \chi_{(-\sigma)(-\sigma')}^{\nu\nu'\omega} $} \\
			\parbox[c][1.0cm]{3cm}{Time reversal} & 
			{$\chi_{\sigma\sigma'}^{\nu\nu'\omega} = \chi_{\sigma'\sigma}^{\nu'\nu\omega}$} \\
			\parbox[c][1.0cm]{3cm}{Particle-hole} & 
			{$\chi_{\sigma\sigma'}^{\nu\nu'\omega} = (\chi_{\sigma\sigma'}^{\nu\nu'\omega})^* $} \\
			\hline
		\end{tabular}
	}
	\label{tab:symmetries}
\end{table}

Table~\ref{tab:symmetries} contains a set of common symmetries that the generalized susceptibility may respect. In the following, we will derive which of these symmetries are needed for $\chi_r^{\nu\nu'\omega}$ to fulfill Eq.~(\ref{equ:chisymmetry}) in the main text. Already from the onset it is clear that the ``swapping'' symmetry, which corresponds to the double application of the crossing symmetry, is of fundamental importance. It corresponds to swapping the particle labels of both the incoming and the outgoing electrons (two swaps), which according to the Pauli principle should leave the state of the system completely unchanged. It is respected by the bare susceptibility as well as the vertical/crossed term given by the third line in Eq.~(\ref{equ:defchi}) by construction, which implies that it is preserved by the BS equations (\ref{equ:defBS}) and its inverse relation in Eq.~(\ref{equ:BSinv}).

The generalized susceptibility in the particle-particle notation only needs to respect spin-SU(2) (su) symmetry, in addition to the swapping symmetry (ss), in order to fulfill Eq.~(\ref{equ:chisymmetry}):
\begin{align}
\chi_{pp,\sigma\sigma'}^{\nu\nu'\omega} &\stackrel{\text{ss}}{=} \chi_{pp,\sigma'\sigma}^{(-\nu-\omega)(-\nu'-\omega)\omega} \nonumber\\
& \stackrel{\text{su}}{=} \chi_{pp,\sigma\sigma'}^{(-\nu-\omega)(-\nu'-\omega)\omega}.
\end{align}
The generalized susceptibility in the particle-hole notation on the other hand does not require spin-SU(2) symmetry to conform with Eq.~(\ref{equ:chisymmetry}), but instead requires complex conjugation (cc) and particle-hole (ph) symmetry,
\begin{align}
\chi_{ph,\sigma\sigma'}^{\nu\nu'\omega} &\stackrel{\text{ss}}{=} \chi_{ph,\sigma'\sigma}^{(\nu'+\omega)(\nu+\omega)(-\omega)} \nonumber\\
& \stackrel{\text{cc}}{=} (\chi_{ph,\sigma\sigma'}^{(-\nu-\omega)(-\nu'-\omega)\omega})^* \nonumber\\
& \stackrel{\text{ph}}{=} \chi_{ph,\sigma\sigma'}^{(-\nu-\omega)(-\nu'-\omega)\omega}.
\end{align}
To conclude, in order for all the parquet channels to decompose into a symmetric and an antisymmetric part the systems need to be paramagnetic and particle-hole symmetric.

\section{The Matrix \texorpdfstring{$M^{r,\omega}$}{M}}
\label{app:matrixelements}

In this appendix we will give the explicit results for the matrix elements of the matrix $M^{r,\omega}$ defined in Eq.~(\ref{equ:calcq1}) for the density, magnetic, and singlet channels. Inserting the explicit expressions for $b_{0,r}^{\nu\omega}$, $b_{1,r}^{\nu\omega}$ and $b_{2,r}^{\nu\omega}$ into this definition, one obtains for the matrix elements $M_{ij}^{r,\omega}$
\begin{widetext}
	\begin{subequations}
		\label{equ:matrixMdef}
		\begin{align}
		\label{equ:matrixMdef11}
		&M_{11}^{r,\omega}=1+\frac{(\mathcal{B}_1^r)^2}{\mathcal{B}_0^r}(1-C_r^\omega)\frac{1}{\beta}\sum_\nu \frac{[\nu(\nu+\omega)-D_r^\omega]^2}{[\nu^2+\frac{U^2}{4}][(\nu+\omega)^2+\frac{U^2}{4}][\nu(\nu+\omega)-B_r^2]},\\[0.2cm]
		\label{equ:matrixMdef12}
		&M_{12}^{r,\omega}=M_{21}^{r,\omega}=\frac{\mathcal{B}_1^r\mathcal{B}_2^r}{\mathcal{B}_0^r}\frac{U^2}{2}\sqrt{U^2+\omega^2}\frac{1}{\beta}\sum_\nu \frac{\nu(\nu+\omega)-D_r^\omega}{[\nu^2+\frac{U^2}{4}][(\nu+\omega)^2+\frac{U^2}{4}][\nu(\nu+\omega)-B_r^2]},\\[0.2cm]
		\label{equ:matrixMdef22}
		&M_{22}^{r,\omega}=1+\frac{(\mathcal{B}_2^r)^2}{\mathcal{B}_0^r}\frac{U^3}{4}\left(\frac{U^2}{1-C_r^\omega}+\omega^2\right)\frac{1}{\beta}\sum_\nu \frac{1}{[\nu^2+\frac{U^2}{4}][(\nu+\omega)^2+\frac{U^2}{4}][\nu(\nu+\omega)-B_r^2]},
		\end{align}
	\end{subequations}
\end{widetext}
where all constants are defined in Table.~\ref{tab:defprefactors} and Eq.~(\ref{equ:defD}), respectively, and we have used that $\omega^2C_r^\omega\!\equiv\!0$ since $C_r^\omega\!\propto\!\delta_{\omega 0}$. The prefactors $(\mathcal{B}_1^r)^2/\mathcal{B}_0^r$, $\mathcal{B}_1^r\mathcal{B}_2^r/\mathcal{B}_0^r$ and $(\mathcal{B}_1^r)^2/\mathcal{B}_0^r$ evaluate to mere phase factors being $-1$, $i$ and $1$ for the density and singlet ($r\!=\!d,s$) channels while they correspond to $1$, $i$ and $-1$ for the magnetic channel ($r\!=\!m$), respectively. Moreover, as one can infer from the corresponding definitions of $B_r$ and $C_r^\omega$ in Tab.~\ref{tab:defprefactors}, the matrix elements for the density and the singlet channel are entirely equivalent. We can, hence, restrict ourselves to the calculation of the matrix elements $M_{ij}^{r,\omega}$ for the density and the magnetic channel in the following.

The frequency sums in Eqs.~(\ref{equ:matrixMdef}) can be evaluated analytically by means of standard methods. Since either $\omega$ or $C_r^\omega$ is nonzero (but not both at the same time), it is convenient to consider the cases $\omega\!=\!0$ and $\omega\!\ne\!0$ separately, for both the density and the magnetic channel. In spite of these simplifications, the actual explicit calculations are still rather involved and, hence, have been carried out with {\em Mathematica}\cite{Wolfram2015} scripts (see the SM\cite{FootnoteSupplemental2018}). 

\vspace{1cm}

\section{Nondiagonal terms of \texorpdfstring{$\overline{\chi}_r^{\nu\nu'\omega}$}{chiinv}}
\label{app:evalnondiag}

For the calculation of the inverse of the generalized susceptibility in Eq.~(\ref{equ:chibarfinal}), the inverse of the matrix $M^{r,\omega}$ is required, which reads as
\begin{equation}
\label{equ:invM}
(M^{r,\omega})^{-1}=\frac{1}{\det M^{r,\omega}}
\begin{pmatrix}
M_{22}^{r,\omega} & -M_{12}^{r,\omega} \\
-M_{12}^{r,\omega} & M_{11}^{r,\omega}	
\end{pmatrix}
\end{equation}
where $\det M^{r,\omega}\!=\!M_{11}^{r,\omega}M_{22}^{r,\omega}-(M_{12}^{r,\omega})^2$ is explicitly given by
\begin{align}
\label{equ:det}
&\det M^{r,\omega}=\frac{U^2+\omega^2}{4U^2(1+e^{\mp\beta U/2})^{-2}+\omega^2}\times\nonumber\\[0.3cm]&\left(1\pm\frac{U\tan[\frac{\beta}{4}(\sqrt{4B_r^2+\omega^2}+\omega)]}{\sqrt{4B_r^2+\omega^2}}\right).
\end{align}
Here, the plus sign has to be adopted for $r\!=\!d,s$ and the minus sign for $r\!=\!m$. The last contribution in Eq.~(\ref{equ:chibarfinal}) in the main text can be now explicitly written as
\begin{widetext}
	\begin{equation}
	\label{equ:chibarexplicit}
	\frac{1}{4b_{0,r}^{\nu\omega}b_{0,r}^{\nu'\omega}}\sum_{k,l=1}^2b_{k,r}^{\nu\omega}\overline{M}_{kl}^{r,\omega}b_{l,r}^{\nu'\omega}=\frac{1}{4b_{0,r}^{\nu\omega}b_{0,r}^{\nu'\omega}}[b_{1,r}^{\nu\omega}M_{22}^{r,\omega}b_{1,r}^{\nu'\omega}-b_{1,r}^{\nu\omega}M_{12}^{r,\omega}b_{2,r}^{\nu'\omega}-b_{2,r}^{\nu\omega}M_{12}^{r,\omega}b_{1,r}^{\nu'\omega}+b_{2,r}^{\nu\omega}M_{11}^{r,\omega}b_{2,r}^{\nu'\omega}]/\det M^{r,\omega},
	\end{equation}
\end{widetext}
where we recall that $\overline{M}^{r,\omega}\!=\!(M^{r,\omega})^{-1}$ [see Eq.~(\ref{equ:chibarfinal})]. Although the actual expressions for the matrix elements $M^{r,\omega}$ are rather complicated, some simplifications are possible for Eq.~(\ref{equ:chibarexplicit}). First, we note that the denominator $(\nu^2\!+\!\frac{U^2}{4})[(\nu\!+\!\omega)^2\!+\!\frac{U^2}{4}]$ appears in $b_{0,r}^{\nu\omega}$ as well as in $b_{i,r}^{\nu\omega}$, $i\!=\!1,2$, and, hence, cancels. This observations suggests the following definitions:
\begin{subequations}
	\label{equ:redefchisplit}
	\begin{align}
	\label{equ:redefchisplit2}
	\tilde{b}_{0,r}^{\nu\omega}&=\mathcal{B}_0^r\frac{\beta}{2}[\nu(\nu+\omega)-B_r^2],\\[0.15cm]
	\label{equ:redefchisplit3}
	\tilde{b}_{1,r}^{\nu\omega}&=\mathcal{B}_1^r\sqrt{U(1-C_r^\omega)}[\nu(\nu+\omega)-D_r^\omega]\nonumber\\&=\mathcal{B}_1^r\sqrt{U(1-C_r^\omega)}\left[\frac{2}{\beta\mathcal{B}_0^r}\tilde{b}_{0,r}^{\nu\omega}+(B_r^2-D_r)\right],\\[0.15cm]
	\label{equ:redefchisplit4}
	\tilde{b}_{2,r}^{\nu\omega}&=\mathcal{B}_2^r\sqrt{\frac{U^3}{4}}\sqrt{\frac{U^2}{1-C_r^\omega}+\omega^2}.
	\end{align}
\end{subequations}
We can now replace $b_{0,r}^{\nu\omega}$, $b_{1,r}^{\nu\omega}$ and $b_{2,r}^{\nu\omega}$ in Eq.~(\ref{equ:chibarexplicit}) by $\tilde{b}_{0,r}^{\nu\omega}$, $\tilde{b}_{1,r}^{\nu\omega}$ and $\tilde{b}_{2,r}^{\nu\omega}$, respectively. Moreover, in Eq.~(\ref{equ:redefchisplit3}) we have expressed $\tilde{b}_{1,r}^{\nu\omega}$ in terms of $\tilde{b}_{0,r}^{\nu\omega}$ plus a constant. This allows us to split the terms in Eq.~(\ref{equ:chibarexplicit}) into three groups: (i) The first group contains all terms where both, $\tilde{b}_{0,r}^{\nu\omega}$ and $\tilde{b}_{0,r}^{\nu'\omega}$, are canceled. These terms do not depend on the fermionic Matsubara frequencies $\nu$ and $\nu'$ (but only on the bosonic transfer frequency $\omega$). (ii) The second group of terms includes contributions which are proportional to either $1/\tilde{b}_{0,r}^{\nu\omega}$ or $1/\tilde{b}_{0,r}^{\nu'\omega}$, i.e., the fermionic frequency dependence of such contributions is given by $1/[\nu(\nu+\omega)-B_r^2]$ or $1/[\nu'(\nu'+\omega)-B_r^2]$, respectively. (iii) The third class of terms is proportional to $1/[\tilde{b}_{0,r}^{\nu\omega}\tilde{b}_{0,r}^{\nu'\omega}]$ and, hence, its frequency dependence is given by $1/([\nu(\nu+\omega)-B_r^2][\nu'(\nu'+\omega)-B_r^2])$. More explicitly, we obtain for the three different types of contributions the following expressions
\begin{widetext}
	{\small
		\begin{subequations}
			\label{equ:nondiagsplitexpl}
			\begin{align}
			\label{equ:nondiagsplitexpl1}
			\frac{1}{\beta^2}\frac{(\mathcal{B}_1^r)^2}{(\mathcal{B}_0^r)^2}&\frac{U(1-C_r^\omega)M_{22}^{r,\omega}}{\det M^{r,\omega}},\\[0.15cm]
			\label{equ:nondiagsplitexpl2}
			\frac{1}{\beta^2}\frac{U\mathcal{B}_1^r}{(\mathcal{B}_0^r)^2}&\left[M_{22}^{r,\omega}(1-C_r^\omega)(B_r^2-D_r^\omega)-M_{12}^{r,\omega}\mathcal{B}_2^r\frac{U}{2}\sqrt{U^2+\omega^2}\right]\frac{1}{\det M^{r,\omega}}\left[\frac{1}{\nu(\nu+\omega)-B_r^2}+\frac{1}{\nu'(\nu'+\omega)-B_r^2}\right],\\[0.15cm]
			\label{equ:nondiagsplitexpl3}
			\frac{1}{\beta^2}\frac{U}{(\mathcal{B}_0^r)^2}&\left[M_{22}^{r,\omega}(\mathcal{B}_r^1)^2(1-C_r^\omega)(B_r^2-D_r^\omega)^2-M_{12}^{r,\omega}\mathcal{B}_1^r\mathcal{B}_2^rU^2\sqrt{U^2+\omega^2}(B_r^2-D_r^\omega)+M_{11}^{r,\omega}(\mathcal{B}_2^r)^2\frac{U^3}{4}\left(\frac{U^2}{1-C_r^\omega}+\omega^2\right)\right]\frac{1}{\det M^{r,\omega}}\nonumber\\&\times\frac{1}{\nu(\nu+\omega)-B_r^2}\frac{1}{\nu'(\nu'+\omega)-B_r^2}.
			\end{align}
		\end{subequations}
	}
\end{widetext}
The corresponding prefactors (in front of the $\nu,\nu'$ dependent contributions) have been evaluated with {\em Mathematica} (see SM\cite{FootnoteSupplemental2018}). For the $\nu,\nu'$ independent term (\ref{equ:nondiagsplitexpl1}) this simplifies to $U(\mathcal{B}_1^r)^2/[(\mathcal{B}_0^r)^2\beta^2]$ which corresponds to the constant $U$ term present in the irreducible vertex functions in the density, magnetic and singlet channels (with the corresponding channel dependent numerical prefactor). The prefactor in Eq.~(\ref{equ:nondiagsplitexpl2}) vanishes and, hence, there are no terms depending either on $\nu$ or $\nu'$ only. Finally, the $\nu,\nu'$ independent prefactor in Eq.~(\ref{equ:nondiagsplitexpl3}) evaluates to a finite value giving rise to a term in $\Gamma_r$ which factorizes with respect to the fermionic Matsubara frequencies $\nu$ and $\nu'$. The explicit result is given in the final expression for $\Gamma_r^{\nu\nu'\omega}$ in Eq.~(\ref{equ:gammafinal}) in the main text.

\vspace{1cm}

\section{The matrix \texorpdfstring{$L^{r,\omega}(\lambda_{r,\text{S}}^\omega)$}{L}}
\label{app:expliciteigen}

In this section we provide more details about the analytical calculation of the matrix $L_{kl}^{r,\omega}(\lambda_{r,\text{S}}^{\omega})$ in Eq.~(\ref{equ:eigenvectorsymmetric2}) for $\lambda_{r,\text{S}}^\omega\!\ne\!0$. The explicit expressions for the corresponding matrix elements are similar to $M_{kl}^{r,\omega}$, with the only difference of an additional $\lambda_{r,\text{S}}^{\omega}$ in the denominator of the terms inside the frequency sums. More specifically, we have
\begin{widetext}
	\begin{subequations}
		\label{equ:matrixLdef}
		\begin{align}
		\label{equ:matrixLdef11}
		&L_{11}^{r,\omega}(\lambda)=1+\frac{(\mathcal{B}_1^r)^2}{\mathcal{B}_0^r}(1-C_r^\omega)\frac{1}{\beta}\sum_\nu \frac{[\nu(\nu+\omega)-D_r^\omega]^2}{[\nu^2+\frac{U^2}{4}][(\nu+\omega)^2+\frac{U^2}{4}]}\frac{1}{[\nu(\nu+\omega)-B_r^2]-\lambda[\nu^2+\frac{U^2}{4}][(\nu+\omega)^2+\frac{U^2}{4}]},\\[0.2cm]
		\label{equ:matrixLdef12}
		&L_{12}^{r,\omega}(\lambda)=L_{21}^{r,\omega}(\lambda)=\frac{\mathcal{B}_1^r\mathcal{B}_2^r}{\mathcal{B}_0^r}\frac{U^2}{2}\sqrt{U^2+\omega^2}\frac{1}{\beta}\sum_\nu \frac{\nu(\nu+\omega)-D_r^\omega}{[\nu^2+\frac{U^2}{4}][(\nu+\omega)^2+\frac{U^2}{4}]}\frac{1}{[\nu(\nu+\omega)-B_r^2]-\lambda[\nu^2+\frac{U^2}{4}][(\nu+\omega)^2+\frac{U^2}{4}]},\\[0.2cm]
		\label{equ:matrixLdef22}
		&L_{22}^{r,\omega}=1+\frac{(\mathcal{B}_2^r)^2}{\mathcal{B}_0^r}\frac{U^3}{4}\left(\frac{U^2}{1-C_r^\omega}+\omega^2\right)\frac{1}{\beta}\sum_\nu \frac{1}{[\nu^2+\frac{U^2}{4}][(\nu+\omega)^2+\frac{U^2}{4}]}\frac{1}{[\nu(\nu+\omega)-B_r^2]-\lambda[\nu^2+\frac{U^2}{4}][(\nu+\omega)^2+\frac{U^2}{4}]},
		\end{align}
	\end{subequations}
	where we have used the definition $\lambda\!\equiv\!\lambda_{r,\text{S}}^{\omega}/(\beta\mathcal{B}_0^r)$ for convenience. In principle, the Matsubara sums in Eqs.~(\ref{equ:matrixLdef}) can be performed analytically. However, in view of their complexity it is advantageous to split up the respective second terms inside the $\nu$-sum (i.e., the ones which contain the $\lambda$) by means of a partial fraction decomposition which gives
	\begin{align}
	\label{equ:partialfraction}
	\frac{1}{[\nu(\nu+\omega)-B_r^2]-\lambda[\nu^2+\frac{U^2}{4}][(\nu+\omega)^2+\frac{U^2}{4}]}&=\frac{-1}{\lambda[\nu(\nu+\omega)-G_{r,+}^{\omega}][\nu(\nu+\omega)-G_{r,-}^{\omega}]}\nonumber\\&=\frac{-1}{\lambda(G_{r,+}^\omega-G_{r,-}^\omega)}\left[\frac{1}{\nu(\nu+\omega)-G_{r,+}^\omega}-\frac{1}{\nu(\nu+\omega)-G_{r,-}^\omega}\right],
	\end{align}
	where the poles $G_{r,\pm}^\omega$ of the denominators are given by
	\begin{equation}
	\label{equ:defpoles}
	G_{r,\pm}^\omega=\frac{U^2}{4}\left[-1+\frac{2}{\lambda U^2}\pm\frac{2}{\lambda U^2}\sqrt{1-\lambda(U^2+4B_r^2)-\lambda^2U^2\omega^2}\right].
	\end{equation}
\end{widetext}
Inserting the partial fraction composition in Eq.~(\ref{equ:partialfraction}) into Eqs.~(\ref{equ:matrixLdef}) one obtains $\nu$-sums which are completely analogous to the ones for $M_{kl}^{r,\omega}$ in Eqs.~(\ref{equ:matrixMdef}) with the only difference that $B_r$ is replaced by $G_{r,\pm}^\omega$. The actual expressions are rather lengthy transcendental functions of $\lambda$ and can be evaluated with {\em Mathematica} with slightly modified versions of the scripts given in the SM\cite{FootnoteSupplemental2018}. The condition for $\lambda$ being an eigenvalue is given by Det$[L_{kl}^{r,\omega}(\lambda)]\!=\!0$. For a vanishing eigenvalue $\lambda\!=\!0$ this relation reduces to Eq.~(\ref{equ:gammaglobdiv}) which defines the $U\beta/2$ values at which a global divergence of $\Gamma_r^{\nu\nu'\omega}$ occurs.

\section{Unphysical behavior of the double occupancy}
\label{app:double}

In this appendix, we show the unphysical behavior of the double occupancy when calculated using one of the unphysical solutions $G_0^{(1)}(\nu)$ or $G_0^{(2)}(\nu)$ discussed in Sec.~\ref{sec:unphys}.

\begin{figure}[t!]
	\centering
	\includegraphics[width=\columnwidth]{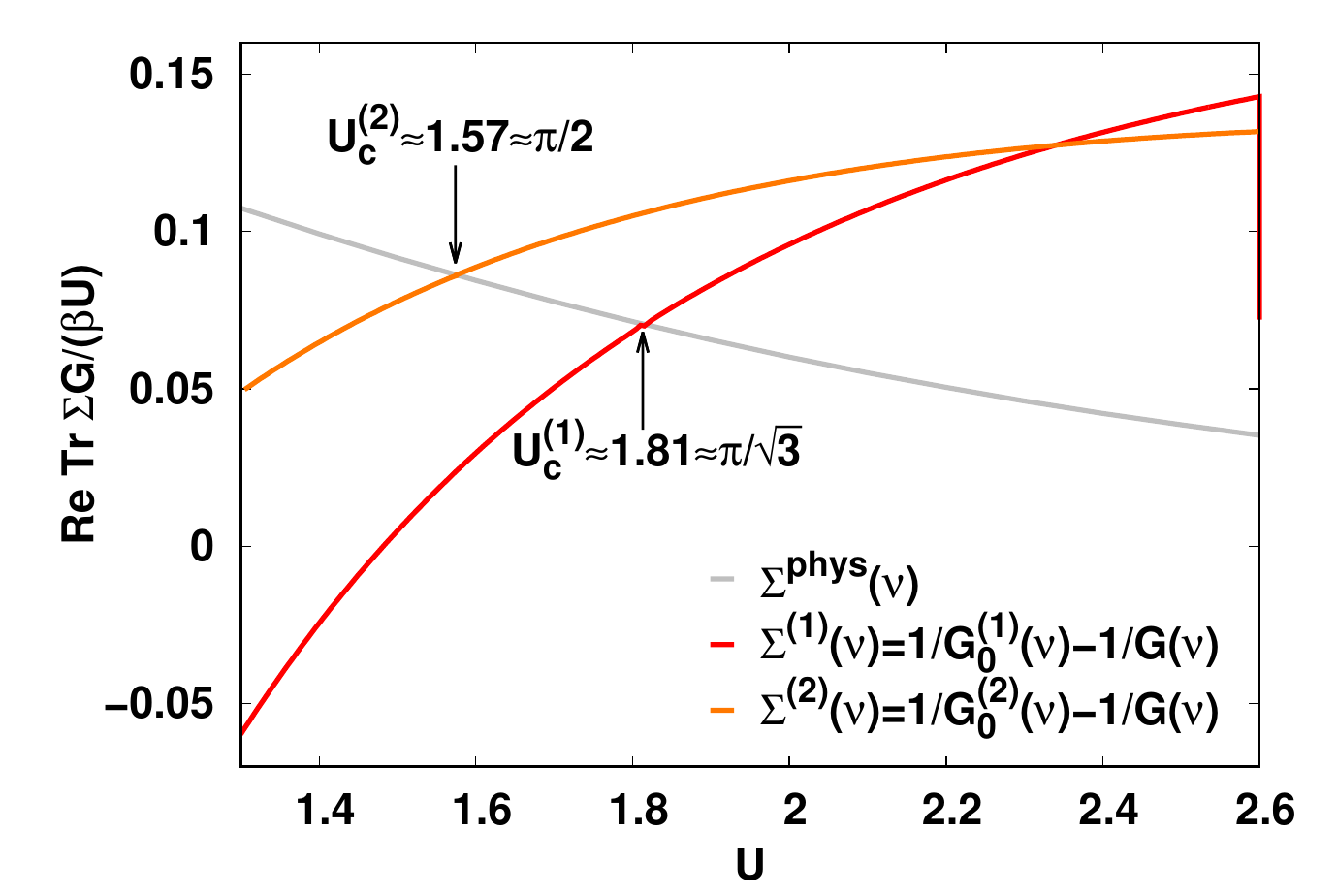}
	\caption{Double occupancy calculated from the Galitskii-Migdal formula\cite{Mahan2000} $\langle\hat{n}_{\uparrow}\hat{n}_{\downarrow}\rangle\!=\!\text{Tr}(\Sigma G)/(U\beta)$ where $\text{Tr}(\Sigma G)\!=\!\sum_{\nu}\Sigma(\nu)G(\nu)$ and $G(\nu)$ is given in Eq.~(\ref{equ:1pgf}). $\beta\!=\!2$. }
	\label{fig:Double_occ}
\end{figure}

The double occupancy has been obtained by the Galitskii-Migdal formula\cite{Mahan2000} where the corresponding unphysical self-energies $\Sigma^{(i)}(\nu)\!=\![G_0^{(i)}(\nu)]^{-1}\!-\!G^{-1}(\nu)$ have been extracted from the corresponding $G_0^{(i)}(\nu)$ [and the physical $G(\nu)$ in Eq.~(\ref{equ:1pgf})] via the Dyson equation~(\ref{equ:dyson}). The results are shown in Fig.~\ref{fig:Double_occ} as a function of $U$. The results obtained from $G_0^{(1)}(\nu)$ (red) and $G_0^{(2)}(\nu)$ (orange) show an unphysical increase of the double occupancy with increasing $U$. Consistent with the analysis in Fig.~\ref{fig:Crossing_G0_orange}, they cross the physical line (gray) at specific values of $U$ where the $G_0^{(i)}(\nu)$ cross $G_0^{\text{phys}}(\nu)$ (see Sec.~\ref{sec:unphys}).

\section{Vertex divergences in the BM in DMFT}
\label{app:inflection}

In this appendix, we will demonstrate that the relation between the (localized) divergence of the irreducible density vertex in the BM ($\Gamma_{d,\text{BM}}^{\nu\nu'\omega}$) and a corresponding minimum in the single-particle Green's function is valid, more generally, for the case of finite bandwidth, when the system is treated in the framework of CPA. To this end we rewrite the single-particle Green's function as
\begin{equation}
\label{def:G}
G(z)=\frac{1}{z-\Sigma(G(z))},
\end{equation}
where $z=i\nu-\Delta(\nu)$ with the hybridization function $\Delta(\nu)$ describing the CPA bath. We have used that for the BM the self-energy $\Sigma$ can be recast as a function of \mbox{$g = G(z)$} by appropriately choosing the {\it physical} solution of the corresponding self-energy functional of the system\cite{Schafer2016}:
\begin{equation}
\label{eq:FKSigmaG}
\Sigma^{\pm}(g) = \frac{\pm \sqrt{1+U^2g^2} - 1}{2g},
\end{equation}
where for a given value of the disorder strength $U$, one has to choose the respective physical branch $\pm$ (for details see Ref.~\onlinecite{Schafer2016}).

The derivative of the Green's function with respect to its frequency argument $\nu$ can be written using the chain rule as
\begin{equation}
\label{eq:dgnu}
\frac{d}{d\nu}G(\nu)=G^{(2)}(i\nu-\Delta(\nu))\left(i-\frac{d}{d\nu}\Delta(\nu)\right)
\end{equation}
where $G^{(2)}(z) \equiv \frac{d }{dz}G(z)$. It is clear that $G(\nu)$ must have an extremal point, irrespective of the lattice structure, when $G^{(2)}(i\nu-\Delta(\nu))$ is zero. 

To evaluate $G^{(2)}$, we first take the {\em ordinary} derivative of Eq.~(\ref{def:G}) and solve for $G^{(2)}(z)$:
\begin{equation}
\label{eq:G2}
G^{(2)}(z) = -\left[g^{-2}- \frac{d}{dg}\Sigma(g) \right]^{-1}_{g=G(z)},
\end{equation}
where the derivative of the self-energy is given by
\begin{equation}
\label{def:gamma2}
\frac{d}{dg}\Sigma(g) = \frac{\sqrt{1+U^2g^2} \mp 1}{2g^2 \sqrt{1+U^2g^2} }.
\end{equation}

The derivative $\frac{d}{dg}\Sigma(G(i\nu-\Delta(\nu)))$ coincides with the diagonal vertex function $\frac{1}{\beta}\Gamma_{d,\text{BM}}^{\nu\nu(\omega=0)} \delta_{\nu\nu'}$ and diverges when
\begin{equation}
\label{eq:FKdivergence}
1+U^2G(i\nu-\Delta(\nu))^2=0.
\end{equation}
Considering $\frac{d}{dg}\Sigma \!\rightarrow\!\infty$ in Eq.~(\ref{eq:G2}) shows that $G^{(2)}$ and thus $\frac{d}{d\nu}G$ goes to zero when $\Gamma_d$ diverges.

Let us finally stress that the above analysis based on ordinary derivatives does not apply to the Hubbard atom. The ordinary derivative $\frac{d}{dg}\Sigma$ cannot even be defined in this case as the self-energy is not just a function of $G(\nu)$, but an (unknown) functional of $G$ evaluated at $\nu$. While $G^{(2)}(z)$ and $\frac{d}{dg}\Sigma$ can be redefined in terms of functional derivatives, the chain rule applied to $\frac{d}{d\nu}G(\nu)$ does not produce these functional derivatives, but only ordinary derivatives.

\end{document}